\documentclass[12pt]{article}

\usepackage{latexsym}
\usepackage{graphicx}
\usepackage{amsmath}
\usepackage{amssymb}
\usepackage{mathrsfs}

\usepackage{psfrag,fancyhdr,epsfig}

\def\inrbra#1{\left\{ #1 \right\}}

\def\beq{\begin{equation}}
\def\eeq{\end{equation}}
\def\bea{\begin{eqnarray}}
\def\eea{\end{eqnarray}}

\begin{document}
\begin{titlepage}

\vspace*{1cm}

\begin{center}

{\bf {\Large {Graviton Emission in the Bulk\\[2mm] by a Simply Rotating
Black Hole}}}
\bigskip \bigskip \medskip

{\bf P. Kanti}$^1$, {\bf H. Kodama}$^2$ {\bf R.A. Konoplya}$^3$, {\bf N. Pappas}$^1$
and {\bf A. Zhidenko}$^4$

\bigskip
$^1${\it Division of Theoretical Physics, Department of Physics,\\
University of Ioannina, Ioannina GR-45110, Greece}

$^2${\it Cosmophysics Group, IPNS, KEK and the Graduate University of Advanced
Studies, 1-1 Oho, Tsukuba 305-0801, Japan}

$^3${\it Department of Physics, Kyoto University, Kyoto 606-8501, Japan}

$^4${\it Instituto de Fisica, Universidade de S\~ao Paulo, C.P. 66318,\\
05315-970, S\~ao Paulo-SP, Brazil}

\bigskip \medskip
{ \bf{Abstract}}
\end{center}
In this work, we study the emission of tensor-type gravitational degrees of freedom from a higher-dimensional, simply rotating black hole in the bulk. The decoupled radial part of the corresponding field equation is first solved analytically in the limit of low-energy emitted particles and low-angular
momentum of the black hole in order to derive the absorption probability. Both the angular and radial equations are then solved numerically, and the comparison of the analytical and numerical results shows a very good agreement in the low and intermediate energy regimes. By using our exact, numerical results we compute the energy and angular-momentum emission rates and their dependence on the spacetime parameters such as the number of additional spacelike dimensions and the angular momentum of the black hole. Particular care is given to the convergence of our results in terms of the number of modes taken into account in the calculation, and the multiplicity of graviton tensor modes that correspond to the same angular-momentum numbers.

\end{titlepage}


\setcounter{page}{1} \noindent

\section{Introduction}

It has been more than a decade since the introduction of the new theories postulating the existence of additional spacelike dimensions in nature; namely, the large extra dimensions \cite{ADD} and warped extra dimensions scenarios \cite{RS} have led to intense research activity of the theoretical as well as the phenomenological consequences of that existence. The introduction of a new, lower than the four-dimensional, fundamental scale for gravity has created the expectation that the elusive quantum theory of gravity might manifest itself soon during high-energy particle collisions, with energies higher than the new gravity scale $M_*$ that may be as low as a few TeVs, at ground-based accelerators. The products of these collisions will inevitably be manifestations of a strong gravity theory, including possibly the creation of higher-dimensional miniature black holes \cite{creation}.

These equally elusive objects, although very small, will be created and decay in a controlled environment and in front of our detectors with the emission of Hawking radiation \cite{hawking} being their most distinctive feature. As a result, there has been a considerable amount of interest in the study of the radiation emission spectra from a higher-dimensional black hole in the literature in recent years (for some reviews, see \cite{Kanti,reviews}). Before reaching the Plank phase, a miniature black hole passes consequently through {\it balding} \cite{braneQNM}, {\it spin-down}, and {\it Schwarzschild} phases \cite{Kanti,reviews}). The starting point was the class of spherically symmetric black holes that were considered to describe the longest and more important {\it Schwarzschild} phase in the life of these objects. The corresponding emission of Hawking radiation was exhaustively studied, both analytically and numerically \cite{KMR1, HK1, Barrau, Jung, BGK, Naylor, Park, Cardoso, CEKT1, Dai}, leading to a number of interesting results.

However, the most generic type of a black hole produced by the collision of two particles with a nonzero impact parameter is a rotating black hole. As a result, the interest was eventually turned to the study of the axially symmetric {\it spin-down} phase \cite{FS-rot, IOP, Nomura, HK2, IOP2, Jung-super, DHKW, CKW, CDKW, CEKT2-3,Chen} that was initially considered to be significantly shorter than and preceding the Schwarzschild phase. However, a recent Monte Carlo simulation \cite{charybdis2} (see also \cite{blackmax}) that has included the effect of rotation of the black hole, has found that the ``spin-down'' phase is not as short-lived as it was thought and that a separate Schwarzschild phase with no angular momentum might not exist at all.

The study of the spin-down phase is important for an additional reason: the question of the energy balance \cite{emparan} between the ``bulk'' and ``brane'' channel during the Hawking radiation has not been answered yet. Studies of the Schwarzschild phase \cite{HK1, Barrau, Naylor, Park, Cardoso, CEKT1, Dai} that included the emission of both scalar fields and gravitons have revealed that the brane channel is in most cases the dominant one, although at certain circumstances the bulk channel can be equally important at specific particle channels. Similar studies have also been performed in more recent years \cite{Jung-rot, CEKT4, kobayashi, CDKW1} for the emission of scalar fields during the spin-down phase, with the dominance of the brane channel still persisting.

In order to give a final answer to the energy balance question, we also need to include in our calculations the emission of gravitons during the spin-down phase. Until recently, the field equations of gravitational perturbations in a higher-dimensional, axially symmetric black-hole background were not known. Even today, we have at our disposal the field equations of specific gravitational modes in certain classes of axially symmetric gravitational backgrounds. The perturbation equations for tensorlike gravitational modes in the case of a higher-dimensional rotating black hole with $D \geq 7$ and equal angular-momentum components was derived in \cite{KLR}. In a subsequent work, one of the authors of this work derived the corresponding equations for tensor-type gravitons for higher-dimensional black holes with one angular-momentum component and $D \geq 7$ again \cite{Kodama-2007}. Then the stability and quasinormal modes of the considered tensor-type gravitational perturbations were investigated in \cite{KKZ2}. Recently, in \cite{MS} perturbation equations were derived for particular scalar, vector and tensor-type gravitational modes for a five-dimensional rotating black hole with two equal angular-momentum components.

In this work, we study the emission of Hawking radiation in the bulk in the form of tensor-type gravitational modes by a higher-dimensional black hole with one angular-momentum component. The geometrical background will be the one considered in \cite{Kodama-2007}, and we will therefore demand the existence of at least three additional spacelike dimensions. Our analysis starts in Sec. 2 where we present in more detail the geometrical setup and theoretical framework. In Sec. 3, we solve analytically the radial part of the graviton field equation in the limit of low-energy emitted particles and low angular momentum of the black hole, and we derive an analytic expression for the absorption probability. We then move, in Sec. 4, to solving the set of angular and radial equations via numerical methods. Our numerical techniques are presented in Sec. 4.1 with our results for the absorption probability described in Sec. 4.2. In the same section, we also perform a comparison of the analytical and numerical results to check the validity of the analytic, approximate method. The exact form of the energy and angular-momentum emission rates in the bulk from the simply rotating black hole in the form of tensor-type gravitons are finally computed and presented in Sec. 4.3. We finish with the presentation of our conclusions in Sec. 5.


\section{Theoretical framework}

As mentioned in the Sec. 1, the perturbation equations for gravitons in a higher-dimensional, rotating black-hole background have been derived in a limited number of cases. Here, we will focus on the case considered in \cite{Kodama-2007} where the higher-dimensional gravitational background can be described by a line element of the form
\begin{equation}
ds^2=G_{MN}\,dz^M\,dz^N=g_{ab}\,dx^a\,dx^b + S^2(x)\,d\Omega_n^2,
\label{D-metric}
\end{equation}
where $\{a,b\}=(0,1,2,3)$ and $d\Omega_n^2$ stands for the line element of an $n$-dimensional unit sphere $S^n$. The above line-element is a special $(4+n)$-dimensional case of a more general class of gravitational backgrounds where the spacetime can be written as the warped product of an $m$-dimensional spacetime ${\cal N}$ and an $n$-dimensional space ${\cal K}$ of constant curvature \cite{KIS-2000, Kodama-Aegean}. In these type of backgrounds, gravitational perturbations can be classified into tensor, vector and scalar types according to their transformation properties as tensors on the constant-curvature spacetime ${\cal K}$. It was this property that allowed for the derivation of perturbation equations for all types of gravitational modes in the case of a maximally symmetric higher-dimensional black-hole background \cite{KI-2003} where $m=2$ and ${\cal K}=S^n$.

The Myers-Perry solution \cite{MP} that describes a $D$-dimensional black hole with $N=[(D-1)/2]$ independent angular-momentum parameters does not, in general, belong to the aforementioned class of line elements. However, when the black hole rotates only in a single two-plane along the four-dimensional spacetime, its line element takes the well-known form
\begin{eqnarray}
&~& \hspace*{-3cm}ds^2 = -\biggl(1-\frac{\mu}{\Sigma\,r^{n-1}}\biggr) dt^2 -
\frac{2 a \mu \sin^2\theta}{\Sigma\,r^{n-1}}\,dt \, d\varphi
+\frac{\Sigma}{\Delta}\,dr^2 +\Sigma\,d\theta^2 \nonumber \\[2mm]
\hspace*{2cm}
&+& \biggl(r^2+a^2+\frac{a^2 \mu \sin^2\theta}{\Sigma\,r^{n-1}}\biggr)
\sin^2\theta\,d\varphi^2 + r^2 \cos^2\theta\, d\Omega_{n}^2,
\label{rot-metric}
\end{eqnarray}
where
\begin{equation}
\Delta = r^2 + a^2 -\frac{\mu}{r^{n-1}}\,, \qquad
\Sigma=r^2 +a^2\,\cos^2\theta\,,
\label{Delta}
\end{equation}
and where we have set $a_1=a$ and $a_2=a_3=\ldots=a_N=0$. The above assumption is justified by the demand that the black hole is created by the collision of particles propagating on our four-dimensional brane: the colliding particles have a nonzero impact parameter only along the usual 3-space which results in a single angular-momentum component once the black hole is created. Then, the black hole's mass $M_{BH}$ and angular momentum $J$ are related to the parameters $a$ and $\mu$ as follows \cite{MP}:
\begin{equation}
M_{BH}=\frac{(n+2) A_{n+2}}{16 \pi G_D}\,\mu\,,  \qquad
J=\frac{2}{n+2}\,M_{BH}\,a\,, \label{def}
\end{equation}
with $G_D$ being the $(4+n)$-dimensional Newton constant, and $A_{n+2}$ the area of an $(n+2)$-dimensional unit sphere given by $A_{n+2}=2 \pi^{(n+3)/2}/\Gamma[(n+3)/2]$.

The line element (\ref{rot-metric}) is a special case of the class of backgrounds considered in \cite{KIS-2000, Kodama-Aegean}, and more specifically of the class described by Eq. (\ref{D-metric}) with $S(x)=r \cos\theta$. As first stated in \cite{Kodama-2007} and later demonstrated in more detail in \cite{KKZ-2009}, the tensor-type gravitational perturbations for the line element (\ref{rot-metric}) -- which exist only for $n \geq 3$, or $D \geq 7$ \cite{Kodama-2007} -- can be expanded in terms of a basis of transverse and traceless harmonic tensors ${\mathbb T}_{ij}^{(\ell,\alpha)}$ on the unit sphere $S^n$ as follows:
\begin{equation}
\delta G_{ij}=2 S^2(x)\,\sum_{\ell,\alpha}\,H_T^{(\ell,\alpha)}(x)\,
{\mathbb T}_{ij}^{(\ell,\alpha)}(y)\,, \label{harmonics}
\end{equation}
where $\{i,j\}$ refer to the $y$ coordinates along the sphere $S^n$, and ${\mathbb T}_{ij}^{(\ell,\alpha)}$ satisfy the eigenvalue equation
\begin{equation}
[\hat\Delta + \ell(\ell+n-1) -2]\,{\mathbb T}_{ij}^{(\ell,\alpha)}=0.
\label{eq-tensor}
\end{equation}
In the above, $\hat\Delta$ is the Laplace-Beltrami operator on $S^n$, and $\ell=2,3,4,\ldots$ an integer number that labels the corresponding eigenvalues. Finally, $\alpha$ is a label to distinguish harmonic tensors with the same eigenvalue.

Under the expansion (\ref{harmonics}), the $(i,j)$ component of Einstein's equation in vacuum leads to the following second-order hyperbolic equation for the amplitude $H_T^{(\ell,\alpha)}(x)$ \cite{KKZ-2009}
\begin{equation}
-\Box H_T -\frac{n}{r\cos\theta}\,g^{ab}\,\partial_a(r \cos\theta)\,\partial_b H_T+
\frac{\ell(\ell+n-1)}{r^2 \cos^2\theta}\,H_T=0,
\label{eq-HT}
\end{equation}
where $\Box$ is the d'Alembertian operator for the metric $g_{ab}(x)$, and where, for simplicity, we have omitted the labels $\{\ell,\alpha\}$. Under the further factorization
\begin{equation}
H_T(x)=e^{-i\omega t}\,e^{im\varphi}\,R(r)\,Q(\theta)\,,
\end{equation}
the above partial differential equation reduces to a set of radial and angular equation, namely,
\begin{equation}
\frac{1}{r^n}\,\partial_r \left(r^n\Delta\,\partial_r R\right) + \left(\frac{K^2}{\Delta} -
\frac{\ell(\ell+n-1)a^2}{r^2} - \Lambda_{j\ell m}\right)R = 0 \,,\label{eq-radial}
\end{equation}
\begin{equation}
\frac{1}{\sin\theta \cos^n \theta}\,\partial_\theta\left(\sin\theta \cos^n \theta\,
\partial_\theta Q\right) +\left(\omega^2 a^2 \cos^2\theta - \frac{m^2}{\sin^2\theta}
- \frac{\ell(\ell+n-1)}{\cos^2\theta} + E_{j\ell m}\right)Q = 0\,. \label{eq-angular}
\end{equation}
In the above, we have used the definitions
\begin{equation}
K=(r^2+a^2)\,\omega-am\,,  \qquad
\Lambda_{j\ell m}=E_{j\ell m}+a^2\omega^2-2am\omega\,,
\end{equation}
with $E_{j\ell m}$ being the separation constant of the two equations, and $j$ a new quantum number that labels the eigenvalues of the angular function $Q(\theta)$.

In order to obtain the complete solution for the wave function of the tensor-type gravitational perturbations of the background (\ref{rot-metric}) one needs to solve the above set of second-order ordinary differential equations (\ref{eq-radial})-(\ref{eq-angular}) for $R$ and $Q$. The same is, in principle, necessary for the computation of the energy emission rate for Hawking radiation in the form of tensor-type gravitational degrees of freedom: the radial equation will yield the expression for the absorption probability (or graybody factor) for the particular type of particles, with the angular equation providing the value of the separation constant $E_{j \ell m}$ that appears in the former equation. The above task can be performed either analytically, in the low-energy and low-angular-momentum limit, or numerically with no restriction on these two parameters. In the next two sections, we will follow both approaches to fulfill this task.


\section{Analytic solution}

As was noted before \cite{Kodama-2007, KKZ-2009} in the context of more general analyses, when the spacetime background has the form of Eq. (\ref{D-metric}), the tensor-type gravitational perturbations are found to satisfy the same field equations that a massless scalar field obeys in the same background. In the present case, the same result also holds as the set of equations (\ref{eq-radial})-(\ref{eq-angular}) are identical to the ones that follow from the scalar field equation
\beq
\frac{1}{\sqrt{-G}}\,\partial_M\left(\sqrt{-G}\,G^{MN}\partial_N \Phi\right)=0\,,
\eeq
if the following expansion of $\Phi$ in terms of the hyperspherical harmonics
${\mathbb Y}^{(\ell,\alpha)}(y)$ on $S^n$ is used
\beq
\Phi(x,y) = e^{-i\omega t}e^{im\varphi}R(r)\,Q(\theta)\,{\mathbb Y}^{(\ell, \alpha)}(y)\,.
\eeq
Given the different nature of the scalar and gravitational degrees of freedom, the two sets of equations differ only in the allowed values of the angular-momentum number $\ell$: whereas, in the scalar case, it satisfies the constraint $\ell \geq 0$ \cite{IUM-2003}, this changes to $\ell \geq 2$ in the case of gravitons \cite{Kodama-2007}. The decoupled set of equations for a massless scalar field propagating in the higher-dimensional background (\ref{rot-metric}) first appeared in \cite{IUM-2003} and were further used in \cite{CEKT4,CDKW1} for the study of the energy emission rates for Hawking radiation emitted by the simply rotating Myers-Perry black hole in the form of scalar fields in the bulk.

The radial equation (\ref{eq-radial}) was analytically solved for scalar fields propagating in the bulk in \cite{CEKT4} in the low-energy and low-angular-momentum approximation. As the equation for tensor-type gravitons is identical, apart from the allowed range of values for $\ell$, the analytic solution in this case follows along the same lines. For this reason, here we give only a brief account of the analysis and the results obtained for the absorption probability in the analytic approach, results that are compared with the exact numerical ones in the next section.

The analytic approach amounts to finding first the asymptotic solutions near the horizon of the black hole ($r\simeq r_h$), and far away from it ($r \gg r_h$) and matching the two in an intermediate zone, to create an analytical solution for $R(r)$ over the whole radial regime. The black hole's horizon radius $r_h$ follows from the equation $\Delta(r_h)=0$, and is given by the relation $r_h^{n+1}=\mu/(1+a_*^2)$, where $a_* \equiv a/r_h$.

In the near-horizon regime ($r\simeq r_h$), Eq. (\ref{eq-radial}) can be rewritten in the form \cite{CEKT4}
\begin{equation}
f\,(1-f)\,\frac{d^2R}{df^2} + (1-D_*\,f)\,\frac{d R}{df} + \biggl[\,\frac{K^2_*}{A_*^2\,f (1-f)}-\frac{\left[\,\ell(\ell+n-1)a_*^2 +
\Lambda_{j \ell m}\right]\,(1+a_*^2)}{A_*^2\,(1-f)}\,\biggr] R=0\,,
\label{eq:NH-1}
\end{equation}
in terms of the new radial variable \cite{CEKT2-3} $f(r) = \Delta(r)/(r^2+a^2)$ and the quantities
\begin{eqnarray}
A_* = (n+1) + (n-1)a_*^2, \qquad K_* = (1+a_*^2) \omega_* - a_* m\,.
\end{eqnarray}
In the above, we have also defined $\omega_* \equiv \omega r_h$ and $D_* \equiv 1 - 4 a_*^2/A_*^2$. Equation (\ref{eq:NH-1}) can be brought to the form of a hypergeometric differential equation with the general solution \cite{Abramowitz}
\begin{eqnarray}
&& \hspace*{-1cm}R_{NH}(f)=A_-\,f^{\alpha}\,(1-f)^\beta\,F(a,b,c;f)
\nonumber \\[1mm] && \hspace*{2cm} +\,
A_+\,f^{-\alpha}\,(1-f)^\beta\,F(a-c+1,b-c+1,2-c;f)\,,
\label{NH-gen}
\end{eqnarray}
where $A_{\pm}$ are integration constants, and the indices ($a$, $b$, $c$) are defined as $a \equiv \alpha + \beta +D_*-1$, $b \equiv \alpha + \beta$, and $c \equiv 1 + 2 \alpha$. In addition, the parameters $\alpha$ and $\beta$ are given by $\alpha_\pm = \pm i K_*/A_*$ and
\begin{equation}
\beta  = \frac{1}{2}\,\biggl[\,(2-D_*) - \sqrt{(D_*-2)^2 - 4\biggl[\frac{K_*^2 - [\,\ell(\ell+n-1)a_*^2 + \Lambda_{j \ell m}]\,(1+a_*^2)}{A_*^2}\biggr]}\,\,\biggr].
\label{beta}
\end{equation}
Close to the horizon, the general solution (\ref{NH-gen}) can be written as the sum of an incoming and an outgoing plane wave after employing a convenient transformation of the radial variable, namely, $y=r_h (1+a_*^2)\ln(f)/A_*$. If we impose the boundary condition that no outgoing modes exist near the black hole's horizon, we can set either $A_-=0$ or $A_+=0$, depending on the choice for the sign of $\alpha$. The two choices are found to be equivalent, thus we choose $\alpha=\alpha_-$ and $A_+=0$. Then, the near-horizon solution acquires the form
\beq R_{NH}(f)=A_-\,f^{\alpha}\,(1-f)^\beta\,F(a,b,c;f)\,.
\label{NH-final} \eeq


On the other hand, in the far-field regime ($r \gg r_h$), Eq. (\ref{eq-radial}) can easily be brought \cite{CEKT4} into the form of a Bessel differential equation if we make the substitution $R(r) = r^{-\left(\frac{n+1}{2}\right)} \tilde R(r)$ and employ a new radial variable $z=\omega r$,
\beq
\frac{d^2\tilde R}{dz^2}+\frac{1}{z}\frac{d \tilde R}{dz}+\left(1 -\frac{E_{j\ell m}+a^2\omega^2+\left(\frac{n+1}{2}\right)^2}{z^2}\right)\tilde R=0 \,.
\label{FF-eq}
\eeq
If we further define for convenience the quantity $\nu=\sqrt{E_{j\ell m}+a^2\omega^2+\left(\frac{n+1}{2}\right)^2}$, the general solution in the far-field regime may be written as
\beq
R_{FF}(r)=\frac{B_1}{r^{\frac{n+1}{2}}}\,J_{\nu}\,(\omega r) + \frac{B_2}{r^{\frac{n+1}{2}}}\,Y_{\nu}\,(\omega r) \,,
\label{FF}
\eeq
where $J_\nu$ and $Y_\nu$ are the Bessel functions of the first and second kind, respectively.

Before the two asymptotic solutions (\ref{NH-final}) and (\ref{FF}) can be matched, they both need to be expanded for intermediate values of the radial variable. To this end, the hypergeometric function appearing in (\ref{NH-final}) needs also to be shifted so that its argument changes from $f$ to $1-f$ by using a well-known relation \cite{Abramowitz, CEKT4}. Then in the limit $r \gg r_h$ or, equivalently $f \rightarrow 1$, the near-horizon solution takes the ``stretched'' form
\beq R_{NH}(r) \simeq A_1\, r^{\,-(n+1)\,\beta} + A_2\,r^{\,(n+1)\,(\beta + D_*-2)}\,,
\label{NH-stretched}
\eeq
with $A_1$ and $A_2$ defined as
\bea 
A_1 &=& A_-\left[(1+a_*^2)\,r_h^{n+1}\right]^\beta \, \frac{\Gamma(c)\Gamma(c-a-b)}{\Gamma(c-a)\Gamma(c-b)}\,, \nonumber\\[1mm]
A_2 &=& A_- \left[(1+a_*^2)\,r_h^{n+1}\right]^{-(\beta + D_*-2)} \, \frac{\Gamma(c)\Gamma(a+b-c)}{\Gamma(a)\Gamma(b)}\,.
\eea
Similarly, the far-field solution (\ref{FF}) in the limit of $r\rightarrow 0$ takes a similar polynomial form
\beq
R_{FF}(r) \simeq \frac{B_1\left(\frac{\omega r}{2} \right)^\nu}{r^{\frac{n+1}{2}} \,\Gamma(\nu+1)}- \frac{B_2}{\pi \, r^{\frac{n+1}{2}}}\,\frac{\Gamma(\nu)}{\left(\frac{\omega r}{2} \right)^{\nu}} \,.
\label{FF-stretched}
\eeq
For the two stretched solutions to perfectly match, the power coefficients of $r$ need to be the same. It can be easily shown that this is indeed the case in the limit of $a_*<1$ and $\omega_*<1$. Then, by ignoring terms of order $(\omega_*^2, a_*^2, a_*\omega_*)$ or higher in the expressions of $\beta$ and $\nu$, we find that $(n+1)\,\beta \simeq -j$,  $(n+1)\,(\beta+D_*-2) \simeq -(j+n+1)$ and $\nu \simeq j+\frac{n+1}{2}$. By identifying the coefficients of the same powers of $r$, we finally obtain the constraint
\beq
\frac{B_1}{B_2} = -\frac{\left(2/\omega r_h\right)^{2j+n+1} \nu\,\Gamma^2(\nu)\,
\Gamma(\alpha+\beta + D_* -1)\,\Gamma(\alpha+\beta)\, \Gamma(2-2\beta - D_*)}
{\pi\,(1+a_*^2)^\frac{2j+n+1}{n+1}\,\Gamma(2\beta + D_*-2)\,\Gamma(2+\alpha -\beta - D_*)\,
\Gamma(1+\alpha-\beta)} \,, \label{B-eq}
\eeq
that guarantees the existence of a smooth, analytic solution for the radial part of the tensor-type graviton wave function for all $r$, valid for small $a_*$ and $\omega_*$.

Let us, at this point, clarify the expression for the eigenvalue $E_{j \ell m}$ that appears both in $\beta$ and $\nu$. This quantity does not exist in closed form, and can be found either numerically or in terms of a power series in the limit of small $a \omega$. We reserve the use of the first method for the next section - in the context of the present analysis, valid in the low-energy and low-angular-momentum, we may use instead the analytic power series expansion \cite{Berti-2006, Cardoso-2005}
\beq
E_{j \ell m}=\sum_{k=0}^{\infty}\,f_k\,(a \omega)^k\,. \label{Ejlm}
\eeq
For the accuracy of our analysis, we keep terms up to 4th order -- the exact expressions\footnote{For consistency, we should point out that in \cite{Berti-2006} the indices $(j,\ell)$ are interchanged compared to the ones in this work, and the total sign of the $f_2$ coefficient should be reversed due to a typographical error.} of the coefficients $f_k$ can be found in \cite{Berti-2006}. It is only in the expansion of the power coefficients of $r$ in the matching process that all terms beyond the first one are ignored; in this case, $E_{j \ell m} \simeq f_0=j(j+n+1)$, where $j\geq \ell+|m|$ and $\frac{j-(\ell+|m|)}{2} \in \{0,\mathbb{Z}^+\}$.

A quantity that determines, to a great extent, the Hawking radiation emission rate of the black hole is the absorption probability $|{\cal A}_{j \ell m}|^2$ -- or graybody factor, since it is the reason for the deviation of the black-hole spectrum from a pure blackbody one. We may derive it, by expanding the far-field solution (\ref{FF-eq}) for $r\rightarrow \infty$, in which case we obtain
\beq
R_{FF}(r) \simeq \frac{1}{r^\frac{n+2}{2}\sqrt{2\pi\omega}}\left[(B_1+iB_2)\, e^{-i\,\left(\omega r - \frac{\pi}{2}\,\nu - \frac{\pi}{4}\right)} + (B_1-iB_2)\,e^{i\,\left(\omega r - \frac{\pi}{2}\,\nu - \frac{\pi}{4}\right)} \right] \,.
\eeq
The absorption probability is then easily determined via the amplitudes of the outgoing and incoming spherical waves, namely,
\beq \left|{\cal A}_{j \ell m}\right|^2 =
1-\left|{\cal R}_{j\ell m}\right|^2
= 1-\left|\frac{B_1-iB_2}{B_1+iB_2}\right|^2 =
\frac{2i\left(B^*-B\right)}{B B^* + i\left(B^*-B\right)+1}\,,
\label{Absorption} \eeq
where $B \equiv B_1/B_2$ is given by Eq. (\ref{B-eq}). The above result can be used to evaluate the absorption probability for the emission of tensor-type gravitons in the bulk, from a simply rotating black hole, in the low-energy and low-angular-momentum limit.

\section{Numerical analysis}

In this section, we use numerical analysis in order to solve both the angular and radial equations for any value of the energy of the emitted particles and angular momentum of the black hole. We start by presenting the main aspects of our numerical techniques, and then we turn to the derivation of exact numerical results for the graybody factor and energy emission spectrum for tensor-type gravitational modes in the bulk.

\subsection{Numerical techniques}

The angular equation (\ref{eq-angular}), in terms of the new variable $x=\cos(2\theta)$, can be written in the following form
\begin{eqnarray}\label{angular-part}
&~& \hspace*{-1cm}2 (1-x^2)\,Q''(x)+[n-1 - (n+3) x\,]\,Q'(x)\\[2mm]
&~& \hspace*{1cm}+\left(\frac{E_{j \ell m}+a^2\omega^2}{2}+\frac{a^2\omega^2(x-1)}{4}
+\frac{m^2}{x-1}-\frac{\ell (\ell + n-1)}{1 + x}\right)Q(x)=0\,.\nonumber
\end{eqnarray}
The above differential equation has been solved in the literature before in different contexts and forms: for instance, its four-dimensional version (that follows for $\ell=n=0$), in the presence of a positive cosmological constant, was solved in \cite{Suzuki:1998vy}; in \cite{Berti-2006}, the above higher-dimensional version was solved for scalar fields, i.e., for $\ell=0,1,2,\ldots$; finally, for tensor-type gravitons living in a higher-dimensional space but in the presence of a negative cosmological constant, the corresponding equation was numerically solved in \cite{KKZ-2009}. Thus, the numerical analysis demanded for solving Eq. (\ref{angular-part}) for tensor-type gravitons living in a higher-dimensional asymptotically flat spacetime is a simplified case of the one presented in \cite{KKZ-2009} -- we refer the interested reader to that work for more details.

The differential equation (\ref{angular-part}) has three regular singular points, at $x=\pm 1$ and $x=\infty$. The angular function $Q(x)$ can be alternatively written as
\beq
Q(x)=(1-x)^{|m|/2}\,(1+x)^{\ell/2}\,y(z)\,,
\eeq
where $x=2z-1$. Under further expansion of the rescaled function $y(z)$ in terms of an infinite series of Jacobi polynomials, supplemented by regularity conditions at $z=0$ and $z=1$, Eq. (\ref{angular-part}) takes the form of an algebraic equation -- a three-term recurrence relation \cite{KKZ-2009}
\beq
c_{k+1}\,\alpha_k + c_k\,\beta_k + c_{k-1}\,\gamma_k=0
\label{recurrence}
\eeq
for the coefficients $c_k$ appearing in the expansion of $y(z)$. In the above relation, the coefficients $\alpha_k$, $\beta_k$, and $\gamma_k$ are constants depending on the fundamental parameters $(\omega_*, a_*, n)$ of the theory, the quantum numbers $(\ell,m)$, the angular eigenvalue $E_{j \ell m}$, and the new index $k=0,1,2,\ldots$ that labels the power of the expansion.

The solution for the angular function $Q(\theta)$ is of limited physical importance for the calculation of the energy emission spectra for the simply rotating black hole. On the other hand, the computation of the eigenvalue $E_{j \ell m}$, that also appears in the radial equation (\ref{eq-radial}) as a separation constant, is of paramount importance. The value of the separation constant can be obtained by applying the infinite continued fractions method \cite{Leaver-1985}. The continued fraction equation \cite{KKZ-2009} follows from the three-term recurrence relation (\ref{recurrence}) and involves ratios of successive terms of the coefficients $\alpha_k$, $\beta_k$, and $\gamma_k$. This equation can be numerically solved in any desired accuracy for the value of $E_{j \ell m}$, for given values of $\omega_*$, $a_*$, and $n$.

In the case of vanishing angular momentum of the black hole, the value of the separation constant can be found in a closed form by the requirement that the corresponding power series expansion of $y(z)$ with a finite number of terms converges \cite{Leaver-1985}. In that case, we obtain \cite{Berti-2006, KKZ-2009}
\begin{equation}\label{Ejlm-nonrotating}
E_{j \ell m} = (2k+\ell+|m|)(2k+\ell+|m|+n+1) \equiv j(j+n+1)\,,
\end{equation}
where, in the last part of the above equation, we have set $j \equiv 2k +\ell+|m|$. Under this alternative definition, the eigenvalue is labeled by a new quantum number with values $j=2,3,4,\ldots$ and, at its lowest order, it coincides with the one for a $(n+2)$ sphere, in agreement with the discussion below Eq. (\ref{Ejlm}).

If the rotation parameter $a$ of the black hole is nonvanishing, the eigenvalues $E_{j \ell m}$ are in principle noninteger and complex. In that case, we can find the value of $E_{j \ell m}(\omega_*, a_*)$, for any value of $\omega_*$ and $a_*$, by using the following procedure.
\begin{enumerate}
\item We start from the nonrotating black hole and find the exact value of $E_{j \ell m}$, for the corresponding $j$, according to Eq. (\ref{Ejlm-nonrotating}).
\item We increase the rotation parameter by a very small amount and search for the closest to the previously found solution for $E_{j \ell m}$.
\item We repeat the previous step until any required value of $a_*$ is reached and all corresponding values of $E_{j \ell m}$ are found.
\end{enumerate}
By following the aforementioned process, we are able to compute the values of the angular separation constant $E_{j \ell m}$, for any $\omega_*$ and $a_*$, and thus to proceed to the numerical integration of the radial equation (\ref{eq-radial}).

Equation (\ref{eq-radial}) can in turn be rewritten in an alternative form under the redefinition of the radial function $R(r)=r^{-n/2}\,(r^2+a^2)^{-1/2}\,P(r)$ and the employment of the tortoise coordinate defined through the relation $dr_\star=(r^2+a^2)\,dr/\Delta$. The new equation then reads
\begin{equation}\label{wave-like-flat}
\frac{d^2 P(r_\star)}{dr_\star^2}+\left[\left(\omega-\frac{am}{r^2+a^2}\right)^2-
\frac{\Delta}{(r^2+a^2)^2}\,U(r)\right]P(r_\star)=0,
\end{equation}
where
\begin{eqnarray}\label{WKBpot}
U(r)&=& \Lambda_{j\ell m}+\frac{\ell(\ell+n-1)a^2}{r^2}+\Delta \left[\frac{n(n+2)}{4r^2} +\frac{3a^2}{(r^2+a^2)^2}\right]\nonumber \\[1mm]
&+& \left[\frac{(n+1)\mu}{r^{n-1}}-2a^2\right] \left(\frac{n}{2r^2}+\frac{1}{r^2+a^2}\right)\,.
\end{eqnarray}
In this form it is straightforward to derive the asymptotic solutions at the horizon and spatial infinity. First, at the horizon, if we set $r \rightarrow r_h$ and $\Delta \rightarrow 0$, we easily obtain
\beq
P(r_\star)\simeq A_1\,e^{i \tilde \omega r_\star}+A_2\,e^{-i \tilde \omega r_\star}\,,
\eeq
where $A_{1,2}$ are integration constants, and
\beq
\tilde \omega \equiv\omega -m\,\Omega_h=\omega-\frac{am}{a^2+r_h^2}\,,
\label{k-def}
\eeq
with $\Omega_h$ the rotation velocity of the black hole. Since no outgoing wave is allowed to classically exist outside the horizon of the black hole, the physically relevant solution of Eq. (\ref{wave-like-flat}) at the horizon is
\begin{equation}\label{cond-horizon}
P(r)=A_2\,e^{-i \tilde \omega r_\star} \simeq
(r-r_h)^{\displaystyle-i(\omega-m\,\Omega_h)\,
\frac{r_h^2+a^2}{\Delta'(r_h)}}\left(Z_h+{\cal O}(r-r_h)\right)\,,
\end{equation}
where $Z_h$ is a rescaled integration constant. As expected, the near-horizon solution (\ref{NH-final}), derived in Sec. 3, reduces to the same expression if we take the limit $f \rightarrow 0$, expand $\Delta$ in powers of $(r-r_h)$ and redefine the integration constant $A_-$.

For the purpose of our numerical analysis, we introduce close to the horizon the new function
\begin{equation}
z(r)=\left(1-\frac{r_h}{r}\right)^{i\,\tilde \omega\,(r_h^2+a^2)/\Delta'(r_h)}P(r).
\end{equation}
Since $P(r)$ satisfies the asymptotic condition (\ref{cond-horizon}), $z(r)$ is regular at the event horizon. We may also fix the value of any undetermined
integration constant, by setting
\beq
z(r_h)=1. \label{z_h}
\eeq
If we then expand $z(r)$ near the event horizon as
\begin{equation}
z(r)=1+z'(r_h)\left(r-r_h\right)+{\cal O}\left(r-r_h\right)^2\,,
\end{equation}
and substitute into Eq. (\ref{wave-like-flat}), we find the value of $z'(r_h)$ which, together with Eq. (\ref{z_h}), are the boundary conditions for our Eq. (\ref{wave-like-flat}) at the horizon.

Next, at spatial infinity ($r \rightarrow \infty$), the two linearly independent solutions of Eq. (\ref{wave-like-flat}) are
\begin{equation}
P_i(r)\sim e^{-i\omega r}, \qquad P_o(r)\sim e^{i\omega r},\label{gen-infinity}
\end{equation}
which describe the ingoing and outgoing wave, respectively. The functions $P_i(r)$ and $P_o(r)$ can be found analytically as series expansions for large $r$ up to any order.

The numerical integration of Eq. (\ref{wave-like-flat}) then proceeds as follows: with the eigenvalue $E_{j \ell m}$ already numerically known for all values of $\omega_*$ and $a_*$, we start from the horizon, with the values of $z(r_h)$ and $z'(r_h)$ as boundary conditions, and move outwards by using the $NDSolve$ built-in function in \emph{Mathematica\textregistered} for $r_h\leq r \leq r_f$, where $r_f \gg r_h$. After the function $P(r)$ is known numerically, we find a fit of this function by considering the superposition of the two solutions (\ref{gen-infinity}) in some region near $r_f$:
\begin{equation}\label{fit-function}
P(r)=Z_i\,P_i(r)+Z_o\,P_o(r).
\end{equation}
The fitting procedure allows us to find the coefficients $Z_i$ and $Z_o$. In order to check the precision of the coefficients we increase the internal precision of $NDSolve$, the value of $r_f$, and the number of terms in the series expansion for $P_i(r)$ and $P_o(r)$, making sure that the values of $Z_i$ and $Z_o$ do not change within the desired precision. The same shooting procedure, though for different boundary conditions, has been used recently in \cite{KZ2} for analysis of stability of higher-dimensional black holes.

Once this process is completed, the quantity $Z_o/Z_i$ gives the ratio of the amplitudes of the outgoing and ingoing modes at a large distance from the black hole, and the absorption probability follows easily through the relation
$$|{\cal A}_{j\ell m}|^2=1-|{\cal R}_{j\ell m}|^2=1-|Z_o/Z_i|^2.$$


\subsection{Absorption probability}

\begin{figure*}[t]
  \begin{center}
 \mbox{\includegraphics[width = .52\textwidth]{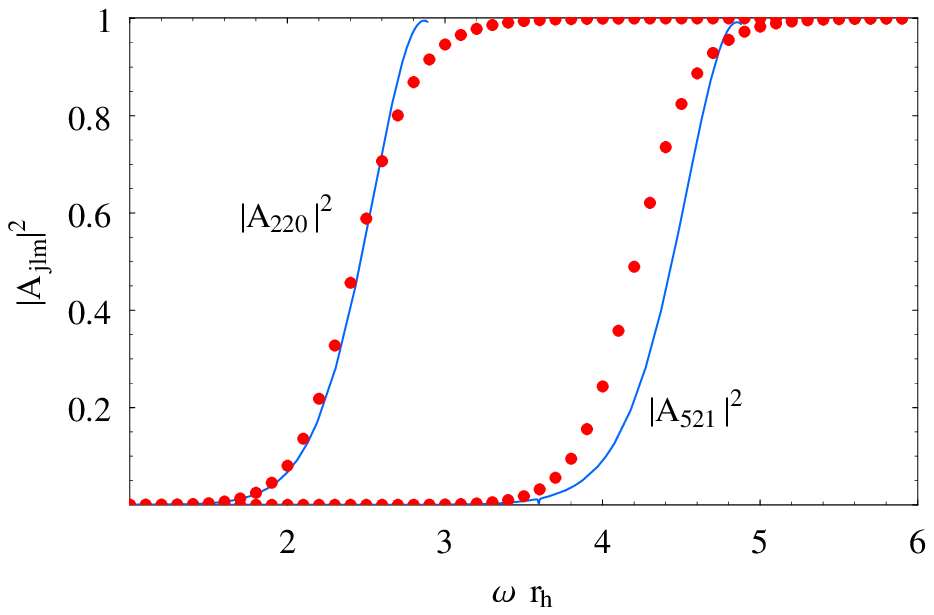}}\hspace*{-0.7cm}
   {\includegraphics[width = .52\textwidth]{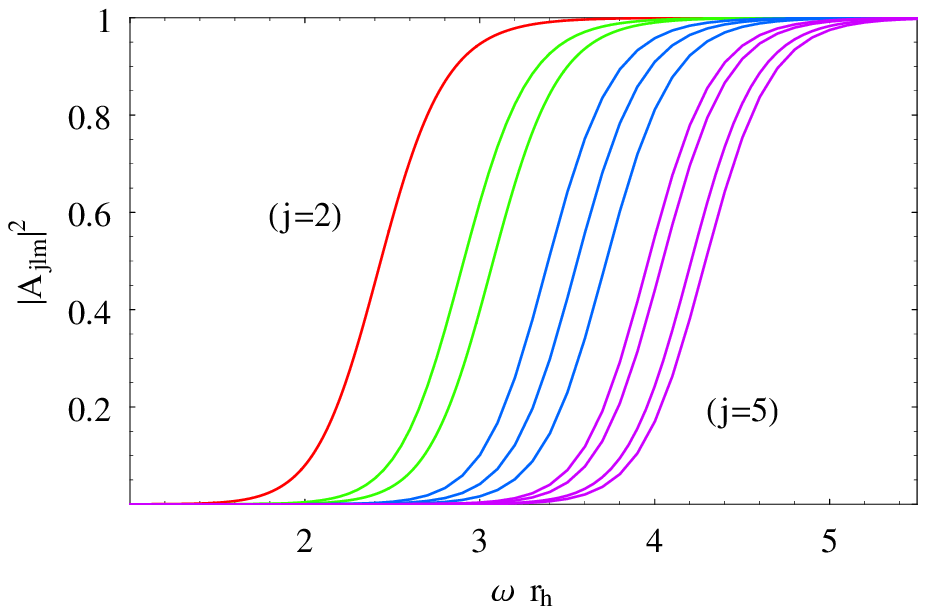}}
  \caption{{\bf (a)} Comparison between our analytical (solid lines) and numerical (data points) results for the graybody factor for the modes $(j=\ell=2,m=0$) and ($j=5,\ell=2,m=1$), for $a= 0.5$ and $D=7$. {\bf (b)} Absorption probabilities for the sets of modes with $\ell=2$, from left to right, $(j=2, m=0)$ (red ine), $(j=3, m=-1,1)$ (green ines), $(j=4, m=-2,0,2)$ (blue lines), and $(j=5, m=-3,-1,1,3)$ (magenta lines), for $a= 0.5$ and $D=7$.}
\label{fig1}
  \end{center}
\end{figure*}

By following the two approaches described in Secs. 3 and 4.1, we have derived analytical approximate results as well as exact numerical ones for the absorption probability for gravitational tensor modes that propagate in the background of a higher-dimensional simply rotating black hole. The two sets of results ought to agree in the low-energy and low-angular-momentum limit, but we expect them to deviate once we move outside these regimes. In order to check the extent of the agreement of the two sets of results as well as its dependence on the particular mode studied, in Fig. \ref{fig1}(a) we depict these two sets for two indicative modes with $(j=2,\ell=2,m=0)$ and $(j=5,\ell=2,m=1)$: the analytical results are given by the solid lines whereas the numerical results are presented as data points - both sets of results correspond to the case with $D=7$ (or $n=3$) and $a=0.5$ (in units of $r_h$). As expected, the agreement between the two sets is indeed very good in the low-energy and even intermediate-energy regime, but inevitably it breaks down as we move towards the high-energy one. The agreement is better for the lowest modes and it worsens for higher modes for which the graybody factor raises to a significant value and approaches unity at an increasingly higher value of the energy parameter $\omega r_h$.

In Fig. \ref{fig1}(b), we examine the aforementioned behavior of the graybody factors for different tensor modes by using exact numerical results. We classify the modes primarily by the angular-momentum number $j$ which can be considered as the total angular-momentum number of the mode, with $\ell$ denoting the angular-momentum along the compact space $S^n$ and $m$ the one in the plane of rotation of the black hole. As in the case of scalar fields \cite{CEKT4, CDKW1}, a set of modes corresponds to each value of $j$: the constraints $j\geq \ell+|m|$ and $\frac{j-(\ell+|m|)}{2} \in\{0,\mathbb{Z}^+\}$ \cite{Berti-2006} dictate that for each value of $j$, $\ell$ can take values in the range $[2,j]$ while, for given $j$ and $\ell$, $m$ can take $j-\ell+1$ values in total. In Fig. \ref{fig1}(b), we display the set of modes corresponding to the values $j=2,3,4,5$ - in order to keep the plot tidy, we fix $\ell=2$ and present the graybody factors for the modes with the $j-\ell+1$ allowed values of $m$ in each case. We may clearly see that as either $j$ or $m$ increases, the corresponding graybody curve shifts to the right and to higher-energies -- a similar behavior would have been observed if we also varied $\ell$.

\begin{figure*}[t]
  \begin{center}
\mbox{\includegraphics[width =.53\textwidth]{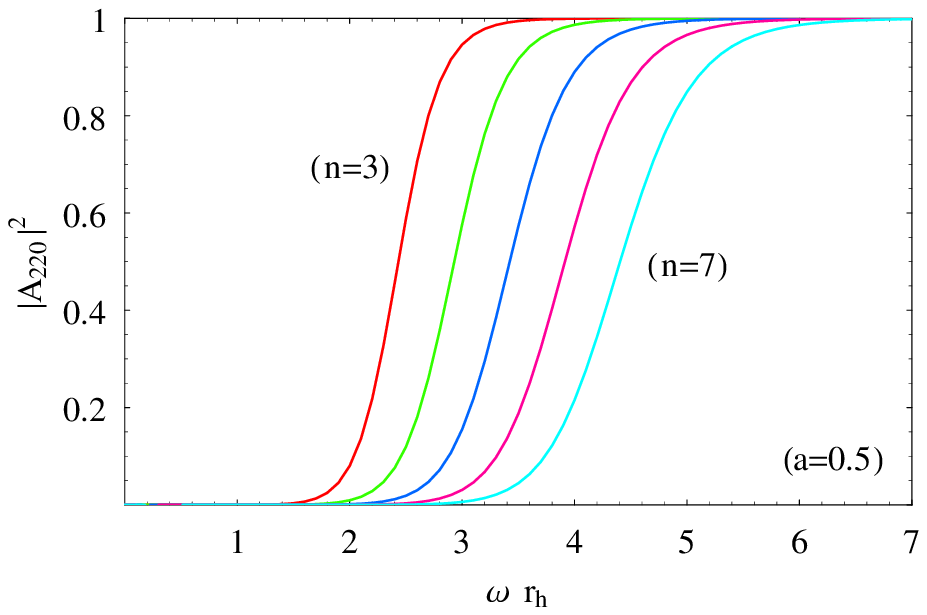}}\hspace*{-0.7cm}
  {\includegraphics[width =.53\textwidth]{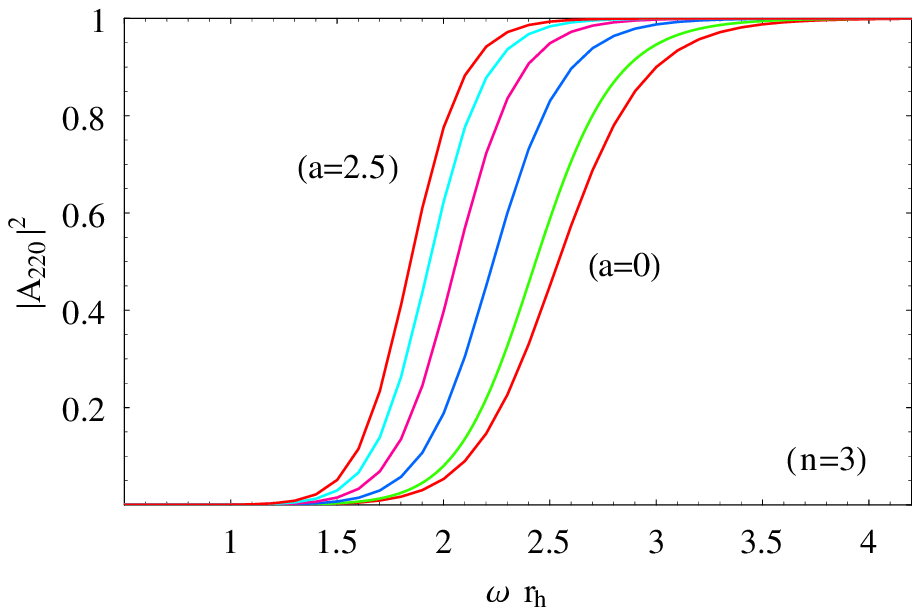}}
    \caption{\label{fig2} Absorption probabilities for the mode $(j=2,\ell=2,m=0$) as
{\bf (a)} a function of $n=3,4,5,6,7$, for $a=0.5$, and {\bf (b)} a function of
$a=0,0.5,1,1.5,2,2.5$, for $n=3$.}
  \end{center}
\end{figure*}

Next, we investigate the dependence of the gravitational tensorial graybody factors on the spacetime parameters of the theory, namely the number of additional spacelike dimensions $n$ and the angular-momentum parameter of the black hole $a$. In Fig. \ref{fig2}(a), we display the absorption probabilities for the indicative mode $(j=2, \ell=2, m=0)$ as $n$ changes from 3 to 7, while keeping the angular-momentum parameter fixed at $a=0.5$. The graybody factors for the gravitational modes in the bulk clearly decrease as the number of transverse-to-the-brane spacelike dimensions increases. For the same mode, in Fig. \ref{fig2}(b), we present the dependence of the graybody factors as $a$ changes from 0 to 2.5, while keeping the dimensionality of spacetime fixed at $D=7$. In this case, the graybody factors for tensorlike gravitons are clearly enhanced as the angular-momentum of the black hole increases. This behavior is in total agreement with the one observed for bulk scalar fields \cite{CEKT4,CDKW1} propagating in the same background.


\subsection{Energy and angular-momentum emission rates}

Having determined the exact value of the absorption probability, we can now proceed to compute the differential emission rates of energy and angular momentum from a higher-dimensional simply rotating black hole in the bulk in the form of tensor-type gravitons. These are given by the following expressions,
\begin{equation}\label{eerate}
\frac{d^2E}{d\omega dt} = \frac{1}{2\pi}\sum_{j,\ell,m}\frac{\omega}
{e^{\tilde \omega/T_H}-1}\,N^\ell_{ST}(S^n)|{\cal A}_{j\ell m}|^2,
\end{equation}
\begin{equation}\label{merate}
\frac{d^2J}{d\omega dt} = \frac{1}{2\pi}\sum_{j,\ell,m}\frac{m}{e^{\tilde \omega/T_H}-1}
N^\ell_{ST}(S^n)|{\cal A}_{j\ell m}|^2,
\end{equation}
where $\tilde \omega$ is defined in Eq. (\ref{k-def}) and the temperature $T_H$ of the black hole is
\begin{equation}
T_{H}= \frac{(n+1)+(n-1)a_*^2}{4\pi (1+a_*^2) r_h}\,.
\label{Temp}
\end{equation}

The quantity $N^\ell_{ST}$ is the multiplicity of the second-rank symmetric, traceless ($T^A{}_A=0$) and divergence-free ($D_BT^{BA}=0$) tensor harmonics $T_{AB}$ that satisfy Eq. (\ref{eq-tensor}). Equivalently, it is the multiplicity of tensor modes on $S^n$ that, under the aforementioned constraints, are described by the same angular-momentum number $\ell$. This number was calculated by Rubin and Ord\'onez in \cite{Rubin.M&Ordonez1984} and found to be
\begin{equation}
N^\ell_{\rm ST}(S^n)
= \frac{(n+1)(n-2)(n+\ell)(l-1)(n+2\ell-1)(n+\ell-3)!}{2(\ell+1)!(n-1)!}\,,
\label{multi}
\end{equation}
for the $\ell$th eigenvalue. The above formula was derived by expanding the tensor harmonics $T_{AB}$ in terms of the harmonic functions $Y^\ell_{(m)}$ and utilizing the representation theory of ${\rm SO}(n+1)$. In the Appendix, we give an alternative proof of the formula \eqref{multi}, in which we start from a similar expansion for the tensor harmonics $T_{AB}$ and determine the multiplicity of modes by finding the number of independent solutions of the equations that the expansion coefficients satisfy -- these equations express the properties of $T_{AB}$ on $S^n$, such as its symmetry, its tracelessness, and transversality. This method is completely elementary and can also be extended into antisymmetric tensor harmonics and higher-rank tensor harmonics.

In order to compute the differential rates (\ref{eerate})-(\ref{merate}), we need to sum the contribution of all tensor modes labeled by the different values of the $(j,\ell,m)$ angular quantum numbers. In practice, the sums need to be truncated at an appropriate high value of each number in such a way that the derived values of the two rates are as close as possible to the real ones. To this end, we adopt the following procedure: we first fix one of the angular numbers and sum over the other two within this range -- in this way we find the contribution of each value of the fixed parameter to the total sum. If the contribution of the highest considered value of an angular number is not small, we increase the particular value range. We repeat the described procedure for all angular numbers until the contribution to the energy and angular momentum emission rates of the highest considered multipole number becomes negligibly small.

\begin{figure*}[t]
  \begin{center}
 \includegraphics[width = .5\textwidth]{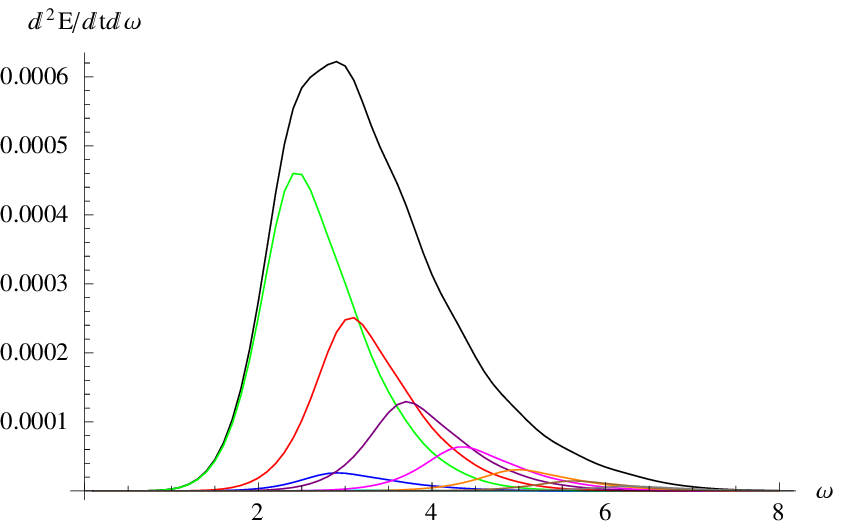}\includegraphics[width = .5\textwidth]{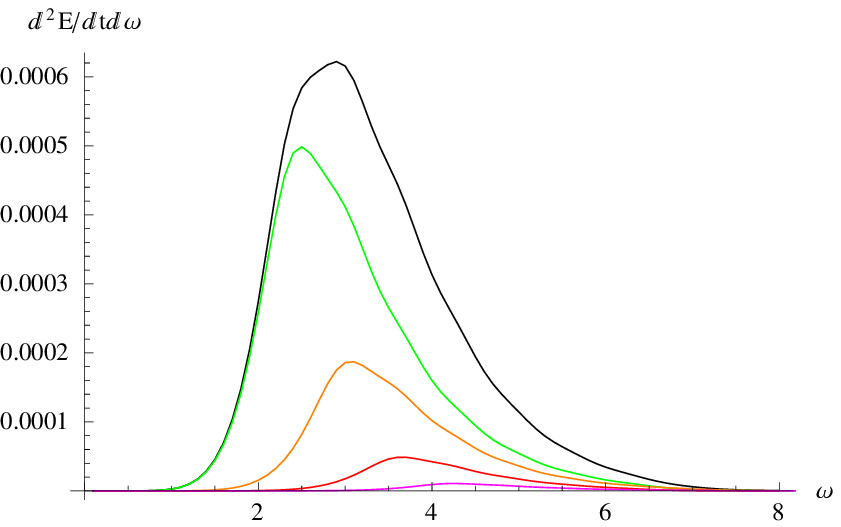}
    \caption{\label{fig3} The energy emission of tensor-type gravitons for $D=7$, $a_*=0.5$ (black line) together with the contributions of different quantum numbers (color lines) are shown. In the left-hand figure: $m=-2$ (cyan line), $m=-1$ (blue line), $m=0$ (green line), $m=1$ (red line), $m=2$ (purple line), $m=3$ (magenta line), $m=4$ (orange line), $m=5$ (brown line), $m=6$ (gray line). In the right-hand figure: $\ell=2$ (green line), $\ell=3$ (orange line), $\ell=4$ (red line), $\ell=5$ (magenta line).\newline The largest contribution correspond to $m=0$. Peaks of positive $m$ contributions ($m=1,2,3,4,5,6$) lay to the right from the peak of $m=0$ contribution.}
  \end{center}
\end{figure*}
\begin{figure*}[t]
  \begin{center}
\includegraphics[width = .5\textwidth]{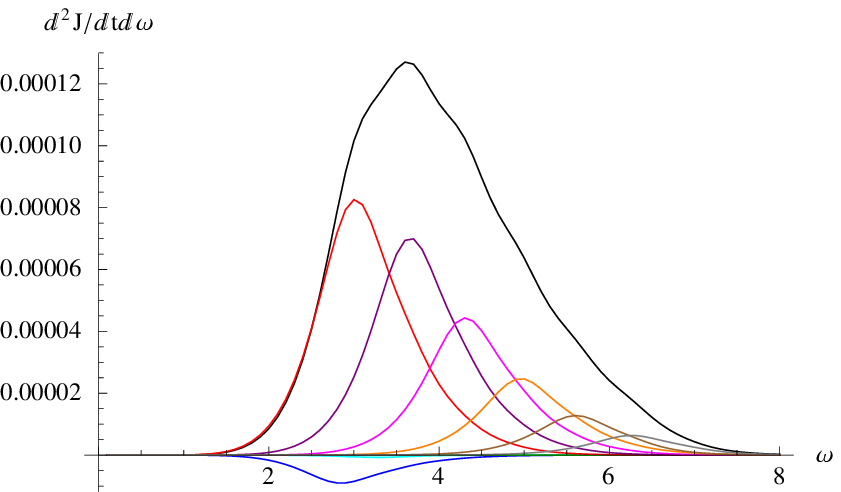}\includegraphics[width = .5\textwidth]{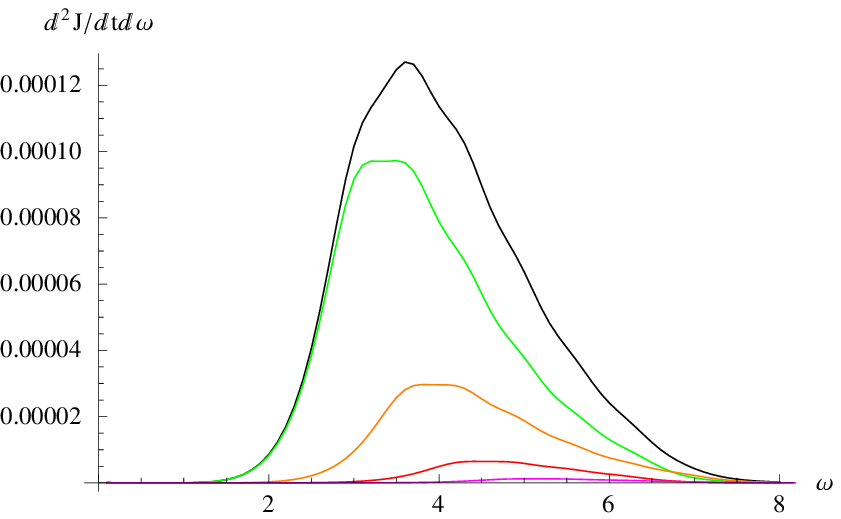}
    \caption{\label{fig4} As above, for the angular-momentum emission of tensor-type gravitons (black line). In the left-hand figure: $m=-2$ (cyan line), $m=-1$ (blue line) $m=0$ (green line), $m=1$ (red line), $m=2$ (purple line), $m=3$ (magenta line), $m=4$ (orange line), $m=5$ (brown line), $m=6$ (gray line). In the right-hand figure (from top to bottom): $\ell=2$ (green line), $\ell=3$ (orange line), $\ell=4$ (red line), $\ell=5$ (magenta line).\newline The largest contribution corresponds to $m=1$. Peaks of other positive m contributions ($m=2,3,4,5,6$) lay to the right from the peak of $m=1$ contribution. The contributions of negative values of m are negative.}	
  \end{center}
\end{figure*}

\begin{figure*}[t]
  \begin{center}
 \includegraphics[width = .7\textwidth]{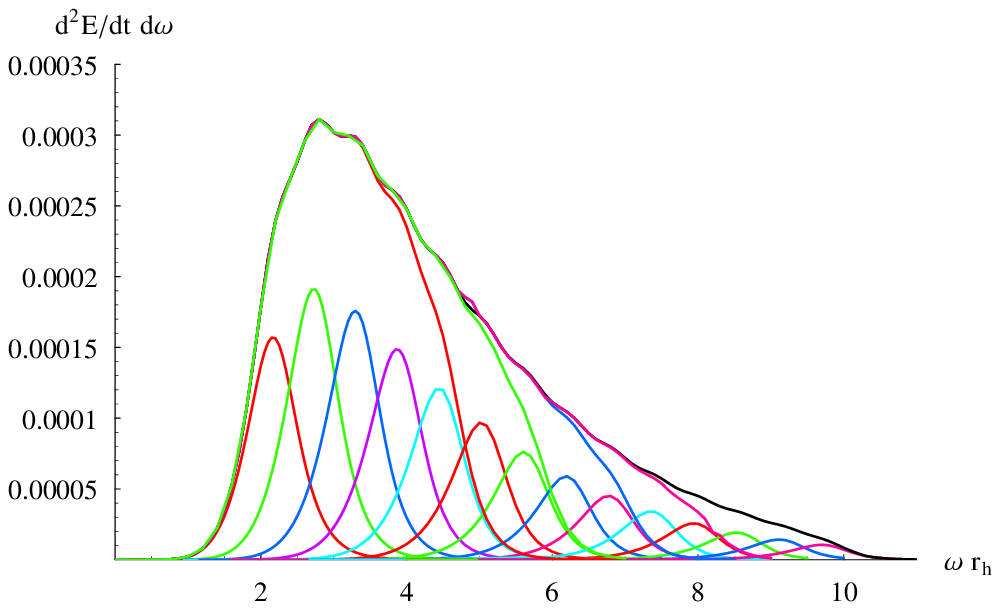}
    \caption{\label{fig5} The energy emission of tensor-type gravitons for $D=7$, $a_*=1$. The different (upper) curves, from left to right, correspond to the highest value of $j$ considered in the sum: $j=6$ (red line), $j=8$ (green line), $j=10$ (blue line), $j=12$ (magenta line), and $j=15$ (black line). }
  \end{center}
\end{figure*}

As an indicative example, in Figs. \ref{fig3}(a,b) and \ref{fig4}(a,b) we display the contributions of the lowest $m$ and $\ell$ tensor modes to the energy and angular-momentum emission rates, respectively, for $D=7$ and $a=0.5$. In all cases, we may observe the increasingly smaller contribution of the higher modes to the specific rate, and thus the convergence of the corresponding sum. This is due to the fact that, according to Fig. \ref{fig1}(b), the higher modes become important at a larger value of the energy parameter $\omega r_h$, and this in practice takes place after the peak of the emission curves -- determined by the temperature of the black hole -- has been reached. As a result the higher modes contribute mostly to the ``tail'' of the emission curves. This is more clearly shown in Fig. \ref{fig5} where the energy emission rate is presented, for $D=7$ and $a=1$, in terms of the contribution of the $j$ modes: as the highest considered value of $j$ increases, from $j=5$ to $j=8$, then to $j=10$, $j=12$, and finally to $j=15$, the emission curve becomes wider and the slope of the tail decreases, whereas the low-energy behavior and the peak of the curve remain unchanged. As mentioned above, in all the cases studied in this work, care was taken so that the change in the emission curves would be negligibly small when a cutoff was imposed on the highest values of all angular numbers. In general, as either $n$ or $a_*$ increases, the number of modes that need to be summed increases, too -- in order to obtain as accurate as possible emission spectra, we have summed up to $j=22$, i.e., $m=20$, in the cases considered.

\begin{figure*}[t]
  \begin{center}
 \includegraphics[width = .45\textwidth]{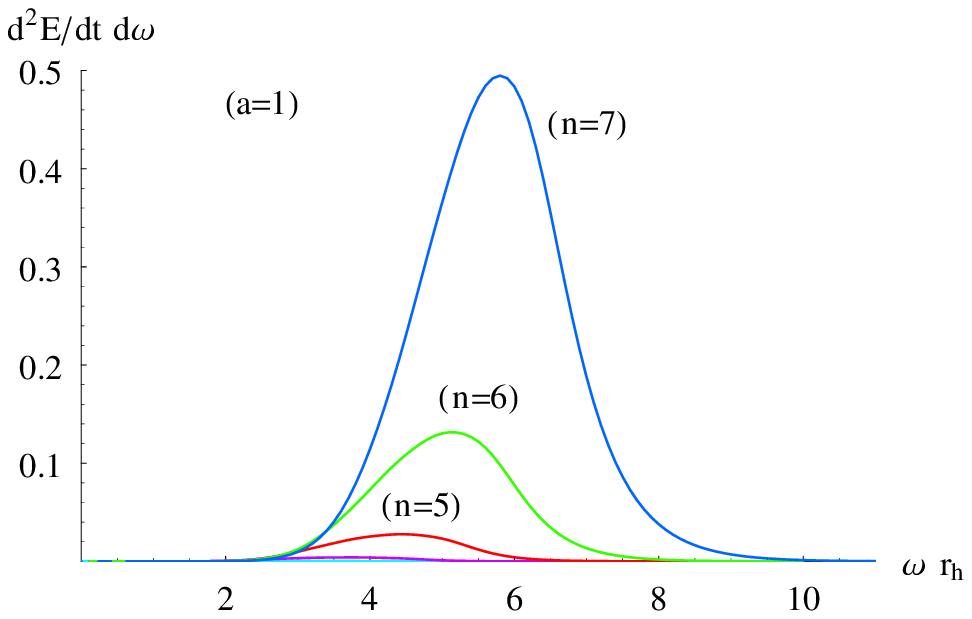}\hspace*{0.5cm}
 \includegraphics[width = .47\textwidth]{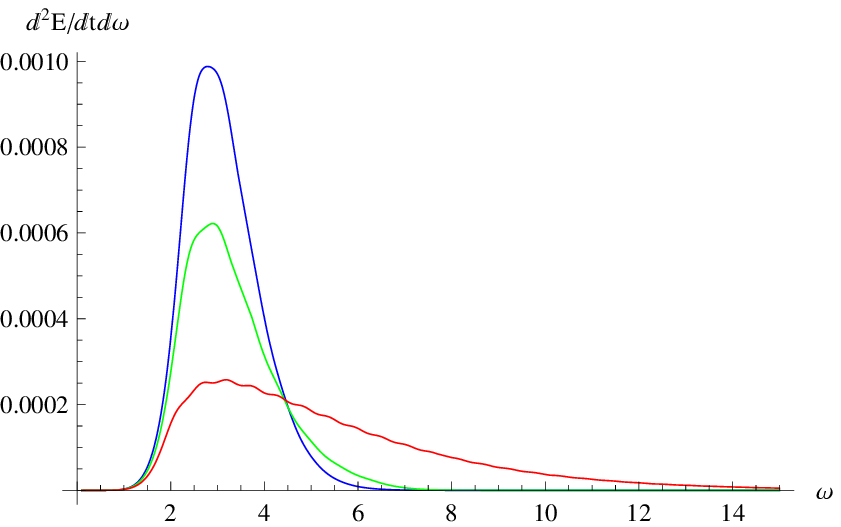}
    \caption{\label{fig6} The energy emission rate for tensor-type gravitons
	in the bulk for {\bf (a)} $a_*=1$ and $n=3,4,5,6,7$ (from bottom to top),
	and for {\bf (b)} $n=3$ and $a_*=0$ (blue line, top), $a_*=0.5$ (green line), $a_*=1.2$ (red line, bottom).}
  \end{center}
\end{figure*}

Next, we turn to the dependence of the energy and angular-momentum emission rates of the black hole on the spacetime parameters, namely, $n$ and $a_*$. In Figs. \ref{fig6}(a,b), we illustrate the dependence of the energy spectrum on the number of additional spacelike dimensions and the angular momentum of the black hole, respectively. As in the case of scalar fields\footnote{As a check of our numerical analysis, we have successfully reproduced the exact results for the energy emission rate of scalar fields in the bulk from a higher-dimensional simply rotating black hole \cite{CDKW1} that were derived with an independent code.} \cite{CEKT4, CDKW1}, the energy emission rate has a very strong dependence on $n$ with an enhancement of almost two orders of magnitude as $n$ changes from $n=3$ to $n=7$. This enhancement is present in all energy regimes with the emission curve becoming significantly taller and wider as $n$ increases. The dependence of the energy spectrum on the angular momentum of the black hole is also nontrivial, although of a smaller magnitude: for the case $n=3$ depicted in Fig. \ref{fig6}(b), the increase of the angular-momentum parameter from $a_*=0$ to $a_*=1.2$ results into the emission of less energy per unit time in the low and intermediate regime and an enhancement in the emission of high-energy modes. As $n$ gets larger, this dependence becomes milder, a feature which is again in accordance with the behavior of the bulk scalar fields emitted by the same black hole spacetime.

\begin{figure*}[t]
  \begin{center}
{\includegraphics[width = .45\textwidth]{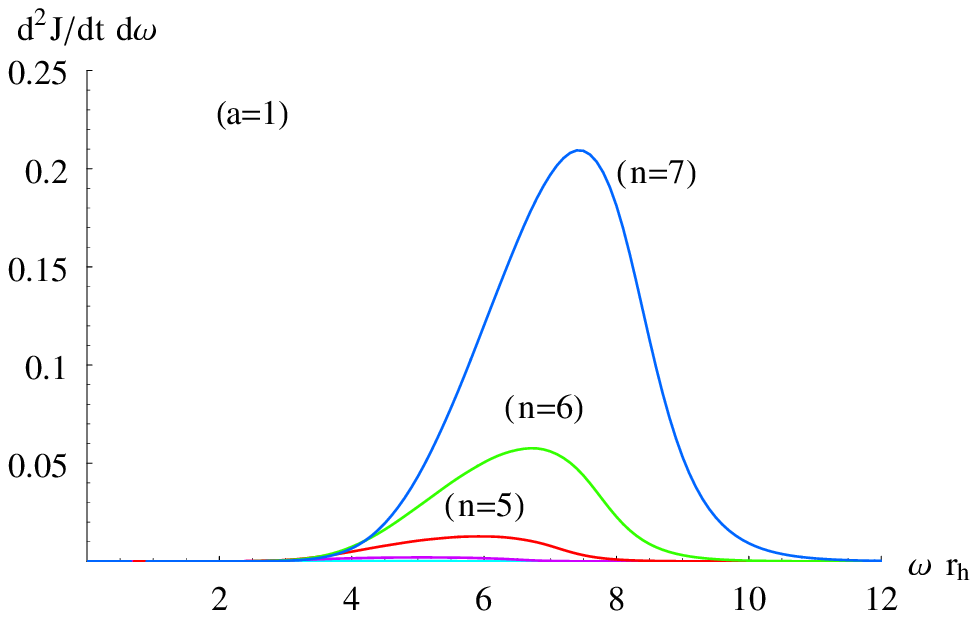}\hspace*{0.5cm}
 \includegraphics[width = .45\textwidth]{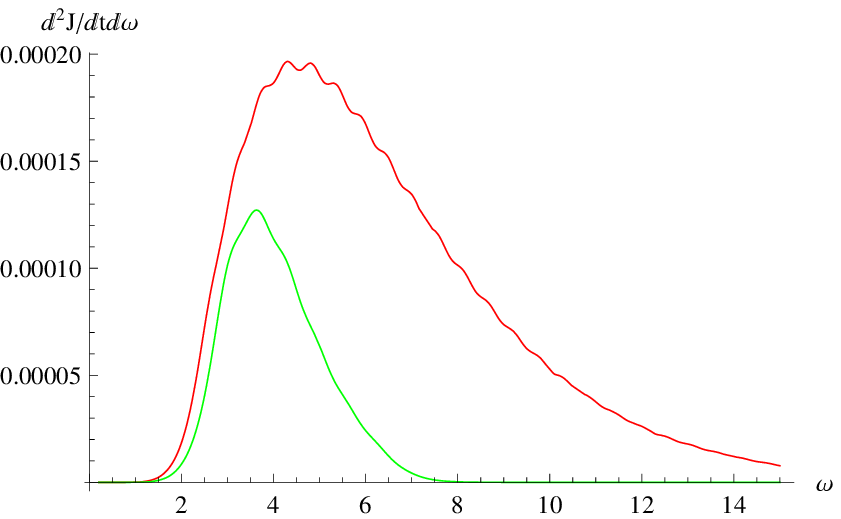}}
    \caption{\label{fig7} The angular-momentum emission rate for tensor-type
	gravitons in the bulk for {\bf (a)} $a_*=1$ and $n=3,4,5,6,7$ (from bottom
	to top), and for {\bf (b)} $n=3$ and $a_*=0.5$ (green line, bottom), $a_*=1.2$ (red line, top).}
  \end{center}
\end{figure*}

In Figs. \ref{fig7}(a,b), we depict the dependence of the angular-momentum emission rate on the same spacelike parameters. As the number of extra dimensions increases, we observe again a significant enhancement in the rate of loss of angular momentum by the black hole. This enhancement reaches more than an order of magnitude and results in the emission of a higher number of modes in all energy regimes. Contrary to what happens in the energy spectrum, the increase in the rotation velocity of the black hole also increases the angular-momentum emission rate from the black hole. The enhancement is significant, although of a smaller magnitude than the one in terms of $n$, leads to the loss of angular momentum via the increased emission of modes in the whole energy spectrum, and manifests itself independently of the dimensionality of spacetime.

\begin{figure*}[t]
  \begin{center}
 \includegraphics[width = .5\textwidth]{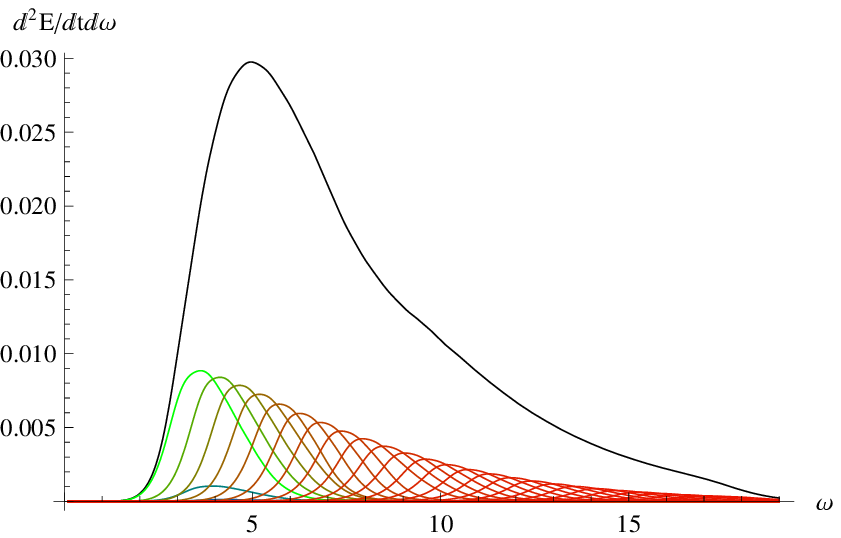}\includegraphics[width = .5\textwidth]{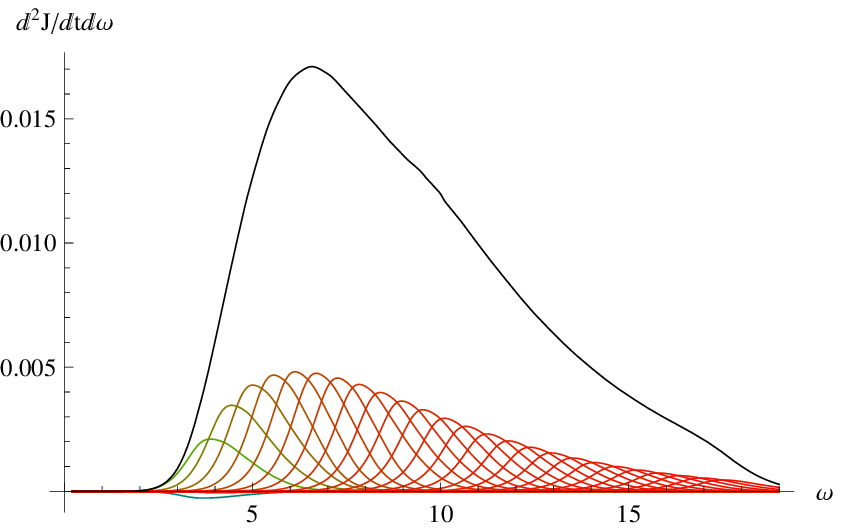}
    \caption{\label{fig8} The energy emission (left) and angular-momentum emission (right) of tensor-type gravitons for $D=9$,
	$a_*=1.2$ (black line) together with the contributions of different quantum numbers $m$ (color lines) are
	shown.}
  \end{center}
\end{figure*}

Finally, in Fig. \ref{fig8} one can see the total energy emission and angular-momentum emission for the tensor-type gravitons for $D=9$ and a fixed $a_*=1.2$ as well as contributions of different quantum numbers $m$, calculated by the accurate shooting method. In Table I the total emission power by the scalar field  for $a_*=1.0$ (taken from \cite{CDKW1}) is given in comparison with the total emission power of tensor-type gravitons. There we can see that although at small number of spacetime dimensions $n$ the contribution of gravitons into the total radiation is tiny, it quickly increases with $n$ and becomes dominant for large $n$.


\begin{table}
\begin{center}
\begin{tabular}{|c|c|c|c|}
\hline
n&Scalar field&Tensor-type gravitons&\\
\hline
3&0.1646&0.0013&0.8\%\\
4&0.3808&0.0222&5.8\%\\
5&0.7709&0.1853&24\%\\
\hline
\end{tabular}
\end{center}
\caption{Total emission power (mass loss rate, in units of $1/r_h^2$) by scalar field ($a_*=1.0$ taken from \cite{CDKW1}) and by tensor-type gravitons ($a_*=1.2$).}\label{comparscalar}
\end{table}

\section{Conclusions}

Because of the complexity of the analysis demanded for the derivation of the perturbed gravitational equations in the background of a higher-dimensional
non-spherically-symmetric black hole, the emission of Hawking radiation in the form of gravitational modes from such a spacetime has been up to now an uncharted territory. In this work, we have used the results of a previous analysis performed by one of the authors \cite{Kodama-2007} according to which the derivation of the field equations for tensor-type gravitational perturbations is indeed possible under the assumption that the spacetime manifold is the warped product of two submanifolds with its line element having the form of Eq. (1). This class of spacetimes includes not only the previously studied higher-dimensional spherically symmetric black-hole backgrounds but also the case of a $(4+n)$-dimensional rotating black hole with a single angular-momentum component along the (3+1)-dimensional brane. In addition, it was further shown that these equations, upon the use of tensor harmonics as a basis, can lead to a set of decoupled ordinary differential equations with respect to the spacetime coordinates.

The derived equations for the tensor-type gravitons propagating in the bulk are found to be identical in form with the ones satisfied by bulk scalar fields. We were thus able to analytically study the problem of the computation of the absorption probability by using techniques employed previously for the emission of scalar fields by the same type of black hole, under proper modifications to allow for the different values of the angular-momentum quantum numbers that characterize the graviton modes. This study led to an analytical expression for the absorption probability for tensor-type bulk gravitons valid in the limit of low-energy emitted modes and low-angular-momentum of the black hole.

In order to derive the complete emission spectra, for arbitrary values of the energy of the emitted mode and angular momentum of the black hole, we also performed an exact numerical analysis to solve both the angular and radial part of the graviton's field equation. In the process, the value of the angular eigenvalue, that appears in and connects the two equations and which does not exist in closed form, was also computed. Having all the above exact results at our disposal, we were thus able to find the value of the absorption probability, or graybody factor, for tensor-type graviton modes in the specific background.

The exact form of the absorption probability in terms of the energy parameter $\omega r_h$ was studied in detail as well as its dependence on the particular graviton mode considered. A comparison between the approximate analytical and exact numerical results for its value was performed, and it was found that, for the lowest graviton modes, the agreement of the two sets of results is remarkably good and extends up to the high-energy regime; as higher modes are considered, the analytical result deviates from the exact one at an increasingly smaller value of the energy. The dependence of the graybody factor on the spacetime parameters, namely the number of additional spacelike dimensions and angular momentum of the black hole, was also investigated. According to our results, the absorption probability for tensor-type gravitons decreases with the number of transverse-to-the-brane dimensions but increases as the black hole rotates faster - this behavior is similar to the one found for bulk scalar fields in previous analyses \cite{CEKT4,CDKW1}.

We next moved to the computation of the emission spectra, namely the energy and angular-momentum ones. The value of the corresponding differential emission rates strongly depends on the number of modes -- characterized by the set of $(j,\ell,m)$ angular-momentum numbers -- that are considered in the sum. For this reason, we performed a careful study of the convergence of our results before imposing a cutoff on the three quantum numbers. In all cases studied, we made sure that the effect of all the higher modes left out of the sum was always negligible. In addition, a technical calculation was performed for the derivation of the multiplicity of tensor modes characterized by the same set of angular-momentum numbers -- this number is distinctly different from the one for bulk scalar fields and affects the value of the differential emission rates.

Combining the above, the energy and angular-momentum emission spectra were finally computed. Both spectra exhibit a very strong dependence on the number of additional spacelike dimensions with the increase in the rate of emission of either energy or angular momentum reaching even 2 or 1 orders of magnitude, respectively. The dependence on the angular momentum of the black hole is more particular: while the angular-momentum emission is clearly enhanced, the differential energy emission rate displays either an enhancement of the high-energy modes and suppression of the other frequencies, for low values of $n$, or a rather mild dependence of the spectrum on $a_*$, for high values of $n$.

Apart from its obvious theoretical interest, the calculation of the emission spectra of a higher-dimensional, simply rotating black hole in the form of gravitons in the bulk has a very important phenomenological interest in the exciting case of the creation of miniature black holes at ground-based accelerators. Previous studies \cite{CEKT4, CDKW1} have revealed that the bulk emission of the other species of particles allowed to propagate in the whole spacetime, namely, the scalar fields, is subdominant compared to the emission that takes place in the form of brane-localized scalar fields. When this is combined with the fact that the total number of fermionic and gauge bosonic degrees of freedom of Standard Model are also restricted, and thus emitted, on the brane, the brane emission channel becomes even more dominant. Addressing the question of energy balance between the brane and bulk channel for the last species, i.e., the gravitons, is of paramount importance for the estimate of the percentage of the total energy of the black hole which is lost in the bulk, and thus of the chances for the potential detection of the produced black hole via the emitted Hawking radiation. The analysis and results presented in this work on the emission of tensor-type graviton modes in the bulk are the first necessary step towards this direction that will hopefully be soon complemented by a similar calculation for the vector and scalar gravitational modes.

\bigskip

{\bf Acknowledgments.}
P.K. and N.P. acknowledge participation in the RTN networks UNIVERSENET-MRTN-CT-2006035863-1 and MRTN-CT-2004-503369.
H. K. was supported in part by Grants-in-Aid for Scientific Research from JSPS (No. 18540265).
R.A.K. was supported by \emph{the Japan Society for the Promotion of Science} (JSPS), Japan.
A.Z. was supported by \emph{Funda\c{c}\~ao de Amparo \`a Pesquisa do Estado de S\~ao Paulo (FAPESP)}, Brazil.

{\bf Note added.} While this manuscript was being written, a complementary work \cite{Doukas} appeared that also studies the emission of tensor-type gravitons
in the bulk by a simply rotating black hole.

\appendix

{\def\CN#1#2{{}_{#1}C_{#2}}
\setcounter{equation}{0}
\renewcommand{\thesection}{\Alph{section}}
\section{Multiplicity Formula for Tensor Harmonics on $S^n$}
\renewcommand{\theequation}{\Alph{section}.\arabic{equation}}

The multiplicity on $S^n$ of the second-rank symmetric tensor harmonics $T_{AB}$, satisfying Eq. (\ref{eq-tensor}) along with the conditions $T^A{}_A=0$ and
$D_BT^{BA}=0$, was calculated by Rubin and Ord\'onez \cite{Rubin.M&Ordonez1984}. They started from the expression for an eigen harmonic tensor
\begin{equation}
T_{i_1i_2}:=T_{AB}D^A\Omega_{i_1}D^B\Omega_{i_2}=\sum_{\ell (m)}b_{i_1i_2}^{\ell(m)}
Y^\ell_{(m)}(\Omega)\,,
\label{Harmonic2Tensor:RO}
\end{equation}
in terms of the harmonic functions $Y^\ell_{(m)}$ with the eigenvalue $\ell(\ell+n-1)$, represented as polynomials of the Cartesian coordinates $\Omega^i$
for the Euclidean space $E^{n+1}$ in which $S^n$ is embedded as $\Omega\cdot\Omega=1$. Here, $D_A$ is the covariant derivative with respect to the standard metric $\gamma_{AB}=\partial_A\Omega\cdot\partial_B\Omega$ of $S^n$. This implies that $T_{i_1i_2}$ transforms as the product of two irreducible representations of ${\rm SO}(n+1)$, $(2,0,\cdots)\times (\ell,0,\cdots)$. Then, they showed that among the irreducible representations obtained by decomposing this product representation, only the $(\ell,2,0,\cdots)$ has the same eigenvalue as $T_{i_1i_2}$ for the standard send-order Casimir operator and its multiplicity is one. Therefore, $N^\ell_{\rm ST}(S^n)$ coincides with the dimension of the irreducible representation $(\ell,2,0,\cdots)$, leading to the formula \eqref{multi}.

In this Appendix, we give an alternative proof of the formula \eqref{multi}, which
as we mentioned in Sec. 4 is more powerful as it can be extended into antisymmetric tensor harmonics and higher-rank tensor harmonics. The starting point is the following expression for $T_{AB}$ similar to \eqref{Harmonic2Tensor:RO}:
\begin{equation}
T_{AB}=a_{j_1\cdots j_\ell ;i_1 i_2}\Omega^{j_1}\cdots\Omega^{j_\ell }
       D_A\Omega^{i_1}D_B\Omega^{i_2},
\label{Harmonic2Tensor:HCR}
\end{equation}
where $a_{j_1\cdots j_\ell ;i_1 i_2}$ is a set of constants satisfying the following conditions:
\begin{subequations} \label{Harmonic2Tensor:HCR:C}
\begin{eqnarray}
&& a_{j_1\cdots j_\ell ;i_1 i_2}=a_{(j_1\cdots j_\ell );i_1 i_2},
\label{Harmonic2Tensor:HCR:C1}\\
&& a_{j_1\cdots j_{\ell -2}j}{}^j{}_{;i_1 i_2}=0,
\label{Harmonic2Tensor:HCR:C2}\\
&& a_{(j_1\cdots j_\ell ;j_{\ell +1})}{}^i
   = 0.
\label{Harmonic2Tensor:HCR:C3}
\end{eqnarray}
\end{subequations}
Utilizing the formula $D_A D_B \Omega^j=-\gamma_{AB}\Omega^j$, it is easy to see that $T_{AB}$ defined by \eqref{Harmonic2Tensor:HCR} satisfies the eigenvalue equation \eqref{eq-tensor}, and the conditions on $a_{*}$  are equivalent to the conditions $T^A{}_A=0$ and $D_BT^{BA}=0$.

What we would like to do is to count the independent number of solutions to the linear algebraic conditions \eqref{Harmonic2Tensor:HCR:C}. The basic idea is to reduce the problem to lower dimensional problems by decomposing the range of the indices $i,j,\cdots=1,\cdots,n+1$ to $\sharp= n+1$ and $i',j',\cdots=1,\cdots,n$.

We start from the condition \eqref{Harmonic2Tensor:HCR:C3}. This condition reads
\begin{equation}
a_{\sharp\cdots\sharp:\sharp\sharp}=0,\quad
a_{\sharp\cdots \sharp j'_1\cdots j'_p:\sharp \sharp}
=-\frac{p}{\ell+1-p}a_{\sharp\cdots \sharp (j'_1\cdots: j'_p)\sharp}
\ (1\le p\le \ell),\quad
a_{(j'_1\cdots j'_\ell:j'_{\ell+1})\sharp}=0,
\end{equation}
for $i=\sharp$ and
\begin{equation}
a_{\sharp\cdots \sharp :\sharp i'}=0,\quad
a_{\sharp\cdots \sharp j'_1\cdots j'_p:\sharp i'}
=-\frac{p}{\ell+1-p}a_{\sharp\cdots \sharp (j'_1\cdots: j'_p)i'}
\ (1\le p\le \ell),\quad
a_{(j'_1\cdots j'_\ell:j'_{\ell+1})i'}=0,
\end{equation}
for $i=i'$. Hence, the condition $a_{(j_1\cdots j_\ell:j_{\ell+1})i}=0$ can be rewritten as
\begin{subequations}
\begin{eqnarray} \label{reduction1}
&& a_{\sharp\cdots \sharp :\sharp \sharp}=0,\
   a_{\sharp\cdots \sharp j' :\sharp \sharp}=0,\
   a_{\sharp\cdots \sharp :\sharp i'}=0,\
   a_{\sharp (j'_1\cdots :j'_\ell j'_{\ell+1})}=0,\
   a_{(j'_1\cdots j'_\ell:j'_{\ell+1})i'}=0,\\
&& a_{\sharp\cdots \sharp j'_1\cdots :j'_p \sharp }
=- \frac{p-1}{\ell+2-p}a_{\sharp\cdots \sharp (j'_1\cdots: j'_{p-1})j'_p}
\quad (2\le p\le \ell+1),\\
&& a_{\sharp\cdots \sharp j'_1\cdots j'_p:\sharp \sharp}
 = \frac{p(p-1)}{(\ell+1-p)(\ell+2-p)}a_{\sharp\cdots \sharp (j'_1\cdots: j'_{p-1}j'_p)}
\quad  (2\le p\le \ell).
\end{eqnarray}
\end{subequations}
%

Next, we consider the condition \eqref{Harmonic2Tensor:HCR:C2}. For $i_1=i'_1$ and $i_2=i'_2$, this condition can be divided into the following set of conditions:
\begin{equation}
a_{\sharp\cdots \sharp j'_1\cdots j'_p:i'_1 i'_2}
   +a_{\sharp\cdots \sharp}{}^{k'}_{k' j'_1\cdots j'_p:i'_1 i'_2}=0
\quad (0\le p \le \ell-2).
\end{equation}
Hence, taking into account the expressions \eqref{reduction1}, we find that all components of $a_{j_1\cdots:i_1i_2}$ can be expressed in terms of $a_{j'_1\cdots j'_\ell:i'_1 i'_2}$ and $a_{\sharp j'_1\cdots j'_{\ell-1}:i'_1 i'_2}$ satisfying the constraints
\begin{equation}
A^{(0)}: a_{(j'_1\cdots j'_\ell:j'_{\ell+1})i'}=0,\quad
B^{(0)}: a_{\sharp (j'_1\cdots: j'_{\ell}j'_{\ell+1})}=0
\label{TH:dConstraint},
\end{equation}
as
\begin{subequations}
\begin{eqnarray} \label{reduction2}
&& \hspace*{-1cm} a_{\sharp\cdots \sharp j'_1\cdots j'_p: i'_1 i'_2}
=(-1)^{[\frac{\ell-p}{2}]}
 a_{(\sharp)}{}^{k'_1}_{k'_1\cdots j'_1\cdots j'_p:i'_1 i'_2}
\qquad (0\le p\le \ell-2),\\
&& \hspace*{-1cm} a_{\sharp\cdots \sharp : i' \sharp }=0,\\
&& \hspace*{-1cm} a_{\sharp\cdots \sharp j'_1\cdots :j'_p \sharp }
= \frac{p-1}{\ell+2-p}(-1)^{[\frac{\ell-p}{2}]}
 a_{(\sharp)}{}^{k'_1}_{k'_1\cdots (j'_1\cdots: j'_{p-1})j'_p}
\qquad (2\le p\le \ell+1),\\
&& \hspace*{-1cm} a_{\sharp\cdots \sharp :\sharp \sharp}=0,\quad
   a_{\sharp\cdots \sharp j' :\sharp \sharp}=0,\\
&& \hspace*{-1cm} a_{\sharp\cdots \sharp j'_1\cdots j'_p:\sharp \sharp}
 =- \frac{p(p-1)}{(\ell+1-p)(\ell+2-p)}(-1)^{[\frac{\ell-p}{2}]}
  a_{(\sharp)}{}^{k'_1}_{k'_1\cdots (j'_1\cdots: j'_{p-1}j'_p)}
\quad (2\le p\le \ell).
\end{eqnarray}
\end{subequations}
%

Similarly, for $i_1=\sharp$ and $i_2=i'$, the condition \eqref{Harmonic2Tensor:HCR:C2}  can be decomposed into the sequence of conditions
\begin{equation}
 a_{\sharp\cdots \sharp j'_1\cdots j'_p:i' \sharp}
   +a_{\sharp\cdots \sharp}{}^{k'}_{k' j'_1\cdots j'_p:i' \sharp}=0
\quad
  (0\le p \le \ell-2),
\end{equation}
which can be written with the help of \eqref{reduction2} as
\begin{equation}
p a_{(\sharp)}{}^{k'_1}_{k'_1\cdots (j'_1\cdots: j'_p)i'}
   +(\ell+1-p) a_{(\sharp)}{}^{k'_1}_{k'_1\cdots j'_1\cdots j'_p}{}^{k'}{}_{:k' i'}=0
\  (0\le p\le \ell-1).
\end{equation}
Because this sequence of equations is consistent with contraction, if they hold for $p=\ell-1,\ell-2$, they hold for the other values of $p$ as well. Further, the equation for $p=\ell-1$ is a contraction of the first of \eqref{TH:dConstraint}. Hence, the only independent condition is
\begin{equation}
(\ell-2)a_{\sharp}{}^{k'}_{k' (j'_1\cdots: j'_{\ell-2})i'}
 + 3a_{\sharp j'_1\cdots j'_{\ell-2}}{}^{k'}{}_{:k' i'}=0.
 \label{TH:dConstraint1}
\end{equation}

Finally, for $i_1=i_2=\sharp$, the condition \eqref{Harmonic2Tensor:HCR:C2} can be decomposed to
\begin{equation}
 a_{\sharp\cdots \sharp j'_1\cdots j'_p:\sharp\sharp}
   +a_{\sharp\cdots \sharp}{}^{k'}_{k' j'_1\cdots j'_p:\sharp \sharp}=0
\quad
  (0\le p \le \ell-2).
\end{equation}
Inserting the relation \eqref{reduction2} into these, we get
\begin{eqnarray*}
&& p(p-1)(\ell-p)(\ell-1-p) a_{(\sharp)}{}^{k'_1}_{k'_1\cdots (j'_1\cdots: j'_{p-1} j'_p)}-
\notag\\
&& (p+1)(p+2)(\ell+1-p)(\ell+2-p)a_{(\sharp)}{}^{k'_1}_{k'_1\cdots (k' k'j'_1\cdots: j'_{p-1} j'_p)}=0
\qquad
 (0\le p\le \ell-2).
\label{TH:dConstriant2}
\end{eqnarray*}
Thus, the number of degrees of freedom of $a_{*}$ is identical to the number of linearly independent solutions to the constraints \eqref{TH:dConstraint}, \eqref{TH:dConstraint1} and \eqref{TH:dConstriant2} for  $a_{\sharp j'_1\cdots: j'_{\ell}j'_{\ell+1}}$ and $a_{j'_1\cdots j'_\ell:j'_{\ell+1}i'}$.

We first calculate the degrees of freedom of $a_{j'_1\cdots j'_\ell:i'_1 i'_2}$. Let us introduce the following symbols representing tensors:
\begin{subequations}
\begin{eqnarray}  \label{XYZW:def}
X^m &:& a^{k'_1}_{k'_1}\cdots ^{k'_m}_{k'_m}{}_{(j'_1\cdots j'_{\ell-2m}: j'_{\ell-2m+1}j'_{\ell-2m+2})}\ (m=0,\cdots,[\ell/2]),\\
Y^m  &:& a^{k'_1}_{k'_1}\cdots ^{k'_{m-1}}_{k'_{m-1}}{}^{k'_m}_{(j'_1\cdots j'_{\ell-2m+1}:j'_{\ell-2m+2}) k'_{m}}\ (m=1,\cdots,[(\ell+1)/2]),\\
Z^m &:& a^{k'_1}_{k'_1}\cdots ^{k'_{m-2}}_{k'_{m-2}}{}^{k'_{m-1}}_{k'_{m-1}}{}_{j'_1\cdots j'_{\ell-2m+2}:}{}^{k'_{m}}_{ k'_{m}}, \ (m=1,\cdots,[\ell/2]+1)\\
W^m &:& a^{k'_1}_{k'_1}\cdots ^{k'_{m-2}}_{k'_{m-2}}{}^{k'_{m-1}k'_m}{}_{j'_1\cdots j'_{\ell-2m+2}:k'_{m-1} k'_{m}}\ (m=2,\cdots,[\ell/2]+1).
\end{eqnarray}
\end{subequations}
In terms of these symbols, the conditions on $a_{j'_1\cdots j'_\ell:i'_1 i'_2}$ are given by $A^{(0)}$ in \eqref{TH:dConstraint} and
\begin{eqnarray}
B^{(m)}&:&  (4m-3)(\ell-2m+2)(\ell-2m+1)X^m
 + 2m(2m-1)\inrbra{2(\ell-2m+2)Y^m+Z^m}=0
\notag\\
&&\quad (m=2,3,\cdots,[\ell/2]).
\end{eqnarray}
First, we show by induction that under the condition $A^{(0)}$, $B^{(m)}$ holds for any $m=2,3,\cdots,[\ell/2]$ if $B^{(2)}$ holds. Let us assume that $B^{(m-1)}$ holds. Then, from the above contraction formulas, the following condition holds:
\begin{eqnarray}
(B^{(m)})' &:& (4m-7)(\ell-2m+2)(\ell-2m+1) X^m+4(2m^2-m-4)(\ell-2m+2)Y^m \notag\\
       && +(4m^2-2m-8)Z^m+8(2m-3)(m-1)W^m=0.
\end{eqnarray}
Further, from the condition $A^{(0)}$, for each level  $m$, we obtain
\begin{subequations}
\begin{eqnarray}
A^{(m)}&:& (\ell-2m+1) X^m + 2m Y^m=0,\\
A_1^{(m)}&:& (\ell-2m+2) Y^m + Z^m + 2(m-1) W^m=0.
\end{eqnarray}
\end{subequations}
It is easy to find that these conditions are related by
\begin{equation}
m (B^{(m)})' -4(2m-3)A_1^{(m)}-2(2m-3)(\ell-2m+2) A^{(m)}= (m-2)B^{(m)}.
\end{equation}
This implies that $B^{(m)}$ holds. Hence, we only have to impose the condition $B^{(2)}$.

Now, we calculate the dimension of the linear set $A^{(0)}$. With the notations $j''_1,\cdots, i''=1,\cdots,n-1$, $A^{(0)}$ is equivalent with
\begin{eqnarray}
&& \hspace*{-1.5cm}
  a_{(j''_1\cdots :j''_{\ell+1}) i''}=0,\quad
  a_{n(j''_1\cdots: j''_\ell j''_{\ell+1})}=0,\\
&&\hspace*{-1.5cm} a_{n\cdots n: nn}=0,\quad
p a_{n\cdots n j''_1\cdots j''_{\ell+1-p}:nn}
 = (\ell-p+1)a_{n\cdots n (j''_1\cdots :j''_{\ell+1-p})n}
\quad (p=1,\dots,\ell),\\
&& \hspace*{-1.5cm} a_{n\cdots n: n i''}=0,\quad
 p a_{n\cdots n j''_1\cdots j''_{\ell+1-p}:n i''}
 = (\ell-p+1)a_{n\cdots n (j''_1\cdots :j''_{\ell+1-p})i''}
\quad (p=1,\dots,\ell).
\end{eqnarray}
Hence, the number $x_n$ of linearly independent equations of $A^{(0)}$ obeys the following recurrence relation:
\begin{equation}
x_n=x_{n-1} + \CN{n+\ell-1}{\ell+1} + n \CN{n+\ell-1}{\ell},\quad
x_1=1
\end{equation}
This can be solved to yield $x_n= n \CN{n+\ell}{\ell+1}$.

Next,  we count the linearly independent equations of $B^{(2)}$. Let us rewrite $B^{(2)}$ as
\begin{subequations}
\begin{eqnarray}
&& 12a_{nn j''_1\cdots j''_{\ell-2}:nn}= \cdots\ (\ell\ge2),\\
&& 36 a_{nnn j''_1\cdots j''_{\ell-3}:nn}=\cdots\ (\ell\ge3),\\
&& (p-1)(5p-6) a_{n\cdots n j''_1\cdots j''_{\ell-p}:nn}=\cdots\ (\ell\ge4,\ p=4,\cdots,\ell),
\end{eqnarray}
\end{subequations}
where the right-hand sides of these are linear combinations of terms of the form $a_{n\cdots n j''_1\cdots j''_{\ell-q}:nn}$ ($q<p$) and $a_{\cdots: j' i''}$. These equations are obviously linearly independent, and the number of them is $\CN{n+\ell-3}{\ell-2}$.
Thus, because we can show that $A^{(0)}$ and $B^{(2)}$ are independent, the number of degrees of freedom of $a_{j'_1\cdots j'_\ell:i'_1 i'_2}$ is given by
\begin{equation}
N_0{}^\ell_n= \frac{n(n+1)}{2}\CN{n+\ell-1}{\ell}- n\CN{n+\ell}{\ell+1} -\CN{n+\ell-3}{\ell-2}.
\end{equation}
%

The degrees of freedom of $a_{\sharp j'_1\cdots j'_{\ell-1}:i'_1 i'_2}$ can be counted in the same way. As in the previous case, let us introduce the symbols $X_\sharp^m$, $Y^m_\sharp$, $Z^m_\sharp$ and $W^m_\sharp$ that are obtained from the definitions \eqref{XYZW:def} by inserting $\sharp$ at the top position of the indices as
$X_\sharp^m : a_\sharp{}^{k'_1}_{k'_1}\cdots ^{k'_m}_{k'_m}{}_{(j'_1\cdots j'_{\ell-2m-1}: j'_{\ell-2m}j'_{\ell-2m+1})}\ (m=0,\cdots,[(\ell-1)/2])$.
Then, in terms of these symbols, the conditions on $a_{\sharp j'_1\cdots j'_{\ell-1}:i'_1 i'_2}$ are given by
\begin{subequations}
\begin{eqnarray}
A^{(1)}_\sharp &:& (\ell-1)a_\sharp^{k'}{}_{k'(j'_1\cdots j'_{\ell-3}:j'_{\ell-2})i'} +2a_\sharp^{k'}{}_{j'_1\cdots j'_{\ell-3}j'_{\ell-2}:k'i'}=0 ,\\
B^{(m)}_\sharp &:& (4m-1)(\ell-2m+1)(\ell-2m)X_\sharp^m
 + 2m(2m+1)\inrbra{2(\ell-2m+1)Y_\sharp^m+Z_\sharp^m}=0
\notag\\
&&\quad (m=0,1,\cdots,[(\ell+1)/2]).
\end{eqnarray}
\end{subequations}

First, we show by induction that under the condition $A_\sharp^{(1)}$, if $B_\sharp^{(0)}$ is satisfied, $B_\sharp^{(m)}$ is also satisfied for any $m=1,3,\cdots,[\ell/2]$. Let us suppose that $B_\sharp^{(m-1)}$ holds. Then, from the contraction rules above, we obtain
\begin{eqnarray}
(B_\sharp^{(m)})' &:& (4m-5)(\ell-2m)(\ell-2m+1) X_\sharp^m+4(2m^2+m-4)(\ell-2m+1)Y_\sharp^m \notag\\
       && +(4m^2+2m-8)Z_\sharp^m+8(2m-1)(m-1)W_\sharp^m.
\end{eqnarray}
Further, applying the contraction rules to the condition $A_\sharp^{(1)}$ leads to the following two equations at each level $m$:
\begin{subequations}
\begin{eqnarray}
A_\sharp^{(m)}&:& (\ell-2m) X_\sharp^m + (2m+1) Y_\sharp^m=0,\\
A_{\sharp 1}^{(m)}&:& (\ell-2m+1) Y_\sharp^m + Z_\sharp^m + (2m-1) W_\sharp^m=0.
\end{eqnarray}
\end{subequations}
We can show that these equations are related by
\begin{equation}
(2m+1) (B_\sharp^{(m)})' -8(m-1)A_{\sharp 1}^{(m)}-8(m-1)(\ell-2m+1) A_\sharp^{(m)}= (2m-3)B_\sharp^{(m)}.
\end{equation}
This implies that $B_\sharp^{(m)}$ holds. Hence, we only have to impose the condition $B_\sharp^{(0)}$.

Next, we count the linearly independent equations of the constraints. First, the condition $A_\sharp^{(1)}$ can be written
\begin{equation}
(p+3)a_{\sharp n  j_1''\cdots j''_{\ell-p}: ni'}=\cdots,\ (p=2,\cdots,\ell),
\end{equation}
where the right-hand side of this equation is a linear combination of terms of the form $a_{\sharp n\cdots n j''_1\cdots j''_{\ell-q}:nn}$ ($q<p$) and  $a_{\cdots: j' i''}$. Hence, the number of linearly independent equations is $n\CN{n+\ell-3}{\ell-2}$.  Next, the number of independent equations of $B^{(0)}_\sharp$ is obviously $\CN{n+\ell}{\ell+1}$. Hence, because we can show that $A^{(1)}_\sharp$ and $B^{(0)}_\sharp$ are independent, the number of the total degrees of freedom of $a_{\sharp j'_1\cdots j'_{\ell-1}:i'_1 i'_2}$ is
\begin{equation}
N_1{}^\ell_n= \frac{n(n+1)}{2}\CN{n+\ell-2}{\ell-1}-\CN{n+1}{\ell+1}
-n\CN{n+\ell-3}{\ell-2}.
\end{equation}
%
Finally, adding the numbers of the degrees of freedom of $a_{j'_1\cdots j'_\ell: i'_1 i'_2}$ and $a_{\sharp j'_1\cdots j'_{\ell-1}:i'_1 i'_2}$, we obtain
\begin{equation}
N^\ell_{ST}(S^n)=\frac{n(n+1)}{2}\left(\CN{n+\ell-1}{\ell}+\CN{n+\ell-2}{\ell-1}\right)
   -(n+1)\left(\CN{n+\ell}{\ell-1}+\CN{n+\ell-3}{\ell-2}\right)\,,
\end{equation}
which after some algebraic manipulation reduces to Eq. (\ref{multi}).
}%


\begin{thebibliography}{99}
\bibitem{ADD} N.~Arkani-Hamed, S.~Dimopoulos and G.~R.~Dvali,
{\it Phys.\ Lett.}\ B {\bf 429}, 263 (1998); 
{\it Phys.\ Rev.}\ D {\bf 59}, 086004 (1999); 
I.~Antoniadis, N.~Arkani-Hamed, S.~Dimopoulos and G.~R.~Dvali,
{\it Phys.\ Lett.}\ B {\bf 436}, 257 (1998).


\bibitem{RS} L. Randall and R. Sundrum, {\it Phys. Rev. Lett.}
{\bf 83} (1999) 3370; {\it Phys. Rev. Lett.} {\bf 83} (1999) 4690.

\bibitem{creation} T.~Banks and W.~Fischler,
hep-th/9906038; \\ 
D.~M.~Eardley and S.~B.~Giddings,
{\it Phys. Rev.} {\bf D 66}, 044011 (2002);\\ 
H.~Yoshino and Y.~Nambu,
{\it Phys. Rev.} {\bf D66}, 065004 (2002); 
{\it ibid.} {\bf D 67}, 024009 (2003);\\
E.~Kohlprath and G.~Veneziano,
{\it JHEP} {\bf 0206}, 057 (2002); \\
V.~Cardoso, O.~J.~C.~Dias and J.~P.~S.~Lemos,
{\it Phys.\ Rev.}\ D {\bf 67}, 064026 (2003);\\
E.~Berti, M.~Cavaglia and L.~Gualtieri,
{\it Phys. Rev.} {\bf D 69}, 124011 (2004);\\ 
S.~B.~Giddings and V.~S.~Rychkov,
{\it Phys.\ Rev.}\ D {\bf 70}, 104026 (2004);\\ 
H.~Yoshino and V.~S.~Rychkov,
{\it Phys.\ Rev.} D {\bf 71} (2005) 104028 ;\\ 
D.~C.~Dai, G.~D.~Starkman and D.~Stojkovic,
{\it Phys.\ Rev.} {\bf 73} (2006) 104037 ;\\
H.~Yoshino and R.~B.~Mann,
{\it Phys.\ Rev.} {\bf 74} (2006) {044003} ;\\
H.~Yoshino, T.~Shiromizu and M.~Shibata,
{\it Phys.\ Rev.} D {\bf 74}, 124022 (2006). 

\bibitem{hawking} S.~W.~Hawking,
{\it Commun.\ Math.\ Phys.}\  {\bf 43}, 199 (1975).

\bibitem{Kanti}
P.~Kanti,
{\it Int. J. Mod. Phys.} A {\bf 19}, 4899 (2004) [hep-ph/0402168];
{\it {Lect.\ Notes Phys.\ }}  {\bf 769}, 387 (2009) [arXiv:0802.2218 [hep-th]];
  arXiv:0903.2147 [hep-th].

\bibitem{reviews}
M.~Cavaglia,
  {\it {Int.\ J.\ Mod.\ Phys.\ }}  A {\bf 18}, 1843 (2003)
  [hep-ph/0210296];
  \\
  G.~L.~Landsberg,
  {\it {Eur.\ Phys.\ J.\ }}  C {\bf 33}, S927 (2004)
  [hep-ex/0310034];
\\
K.~Cheung,
  hep-ph/0409028;
\\
S.~Hossenfelder,
  hep-ph/0412265;
\\
C.~M.~ Harris,
hep-ph/0502005;
\\
A.~S.~Majumdar and N.~Mukherjee,
  {\it {Int.\ J.\ Mod.\ Phys.\ }}  D {\bf 14}, 1095 (2005)
  [astro-ph/0503473];
\\
E.~Winstanley,
  arXiv:0708.2656 [hep-th];
\\
R.~Emparan and H.~S.~Reall,
  Living Rev.\ Rel.\  {\bf 11}, 6 (2008).
  arXiv:0801.3471 [hep-th].


\bibitem{braneQNM}
  E.~Berti, V.~Cardoso and A.~O.~Starinets,
  arXiv:0905.2975 [gr-qc];
  R.~A.~Konoplya,
  Phys.\ Rev.\  D {\bf 68}, 124017 (2003)
  [arXiv:hep-th/0309030];
  P.~Kanti and R.~A.~Konoplya,
  Phys.\ Rev.\  D {\bf 73}, 044002 (2006)
  [arXiv:hep-th/0512257];
  H.~Ishihara, M.~Kimura, R.~A.~Konoplya, K.~Murata, J.~Soda and A.~Zhidenko,
  Phys.\ Rev.\  D {\bf 77}, 084019 (2008)
  [arXiv:0802.0655 [hep-th]];
  E.~Berti, K.~D.~Kokkotas and E.~Papantonopoulos,
  Phys.\ Rev.\  D {\bf 68}, 064020 (2003)
  [arXiv:gr-qc/0306106].



\bibitem{KMR1}
P.~Kanti and J.~March-Russell,
{\it Phys.\ Rev.}\ D {\bf 66}, 024023 (2002);
{\it Phys.\ Rev.}\ D {\bf 67}, 104019 (2003).

\bibitem{HK1}
  C.~M.~Harris and P.~Kanti,
  {\it JHEP} {\bf 0310} (2003) 014.


\bibitem{Barrau} A.~Barrau, J.~Grain and S.~O.~Alexeyev,
{\it Phys. Lett.} B {\bf 584}, 114 (2004);
J.~Grain, A.~Barrau and P.~Kanti,
{\it Phys.\ Rev.} D {\bf 72}, 104016 (2005);
T.~G.~Rizzo,
{\it Class.\ Quant.\ Grav.}  {\bf 23}, 4263 (2006).

\bibitem{Jung}
E.~l.~Jung, S.~H.~Kim and D.~K.~Park,
{\it Phys.\ Lett.} B {\bf 586} (2004) 390;
{\it JHEP} {\bf 0409} (2004) 005;
{\it Phys.\ Lett.}\ B {\bf 602} (2004) 105;
{\it Phys.\ Lett.} B {\bf 614} (2005) 78;
E.~Jung and D.~K.~Park,
{\it Nucl.\ Phys.}\ B {\bf 717} (2005) 272;
hep-th/0506204.

\bibitem{BGK}
P.~Kanti, J.~Grain and A.~Barrau,
{\it Phys.\ Rev.} D {\bf 71} (2005) 104002.

\bibitem{Naylor}
  A.~S.~Cornell, W.~Naylor and M.~Sasaki,
  {\it JHEP} {\bf 0602} (2006) 012.

\bibitem{Park}
  D.~K.~Park,
  {\it Class.\ Quant.\ Grav.}  {\bf 23} (2006) 4101.

\bibitem{Cardoso}
  V.~Cardoso, M.~Cavaglia and L.~Gualtieri,
  {\it Phys.\ Rev.\ Lett.}  {\bf 96}, 071301 (2006)
  [Erratum-ibid.\  {\bf 96}, 219902 (2006)];
  {\it JHEP} {\bf 0602}, 021 (2006).

\bibitem{CEKT1}
  S.~Creek, O.~Efthimiou, P.~Kanti and K.~Tamvakis,
  {\it Phys.\ Lett.} B {\bf 635} (2006) 39;
  O.~Efthimiou,
  hep-th/0609144. 

\bibitem{Dai}
D.~C.~Dai, N.~Kaloper, G.~D.~Starkman and D.~Stojkovic,
{\it Phys.\ Rev.}  D {\bf 75}, 024043 (2007).

\bibitem{FS-rot}
V.~P.~Frolov and D.~Stojkovic,
{\it Phys.\ Rev.}\ D {\bf 67}, 084004 (2003).

\bibitem{IOP}
  D.~Ida, K.~y.~Oda and S.~C.~Park,
  {\it Phys.\ Rev.}  D {\bf 67}, 064025 (2003)
  [Erratum-ibid.\  D {\bf 69}, 049901 (2004)].

\bibitem{Nomura} H.~Nomura, S.~Yoshida, M.~Tanabe and K.~i.~Maeda,
{\it Prog.\ Theor.\ Phys.}  {\bf 114}, 707 (2005).


\bibitem{HK2}
  C.~M.~Harris and P.~Kanti,
  {\it Phys.\ Lett.}  B {\bf 633}, 106 (2006).

\bibitem{IOP2}
D.~Ida, K.~y.~Oda and S.~C.~Park,
{\it Phys.\ Rev.} D {\bf 71}, 124039 (2005);
{\it Phys.\ Rev.} D {\bf 73}, 124022 (2006).

\bibitem{Jung-super} E.~Jung, S.~Kim and D.~K.~Park,
{\it Phys.\ Lett.} B {\bf 615}, 273 (2005);
{\it Phys.\ Lett.} B {\bf 619}, 347 (2005);

\bibitem{DHKW}
  G.~Duffy, C.~Harris, P.~Kanti and E.~Winstanley,
  {\it JHEP} {\bf 0509}, 049 (2005). 

\bibitem{CKW}
M.~Casals, P.~Kanti and E.~Winstanley,
{\it JHEP} {\bf 0602}, 051 (2006). 

\bibitem{CDKW} M.~Casals, S.~R.~Dolan, P.~Kanti and E.~Winstanley,
{\it JHEP} {\bf 0703}, 019 (2007).

\bibitem{CEKT2-3}
  S.~Creek, O.~Efthimiou, P.~Kanti and K.~Tamvakis,
  {\it Phys.\ Rev.}  D {\bf 75} (2007) 084043;
  {\it Phys.\ Rev.}  D {\bf 76} (2007) 104013.

\bibitem{Chen}
S.~Chen, B.~Wang, R.~K.~Su and W.~Y.~Hwang,
  {\it {JHEP}} {\bf 0803}, 019 (2008)
  [arXiv:0711.3599 [hep-th]].

\bibitem{charybdis2}
J.~A.~Frost, J.~R.~Gaunt, M.~O.~P.~Sampaio, M.~Casals, S.~R.~Dolan,
M.~A.~Parker and B.~R.~Webber,
  arXiv:0904.0979 [hep-ph].

\bibitem{blackmax}
D-C.~Dai, G.~Starkman, D.~Stojkovic, C.~Issever, E.~Rizvi,
and J.~Tseng,
{\it Phys.\ Rev. }  D {\bf 77}, 076007 (2008) [arXiv:0711.3012 [hep-ph]].

\bibitem{emparan}
R.~Emparan, G.~T.~Horowitz and R.~C.~Myers,
{\it Phys.\ Rev.\ Lett.}\ {\bf 85}, 499 (2000) [hep-th/0003118].

\bibitem{Jung-rot}
E.~Jung and D.~K.~Park,
{\it Nucl.\ Phys.} B {\bf 731}, 171 (2005).
{\it Mod.\ Phys.\ Lett.\ } A {\bf{22}}, 1635 (2007) [hep-th/0612043].

\bibitem{CEKT4}
S.~Creek, O.~Efthimiou, P.~Kanti and K.~Tamvakis,
{\it Phys. Lett.}\ B {{\bf 656}}, 102 (2007) [arXiv: 0709.0241 [hep-th]].

\bibitem{kobayashi}
T.~Kobayashi, M.~Nozawa, Y.~Takamizu,
{\it {Phys.\ Rev.\ }} D {\bf 77}, 044022 (2008)
[arXiv:0711.1395 [hep-th]].

\bibitem{CDKW1}
M.~Casals, S.~R.~Dolan, P.~Kanti and E.~Winstanley,
  {\it {JHEP}} {\bf 0806}, 071 (2008)
  [arXiv:0801.4910 [hep-th]].

\bibitem{KLR} H.~K.~Kunduri, J.~Lucietti and H.~S.~Reall,
{\it Phys. Rev.} {\bf D74}, 084021 (2006) [hep-th/0606076].

\bibitem{Kodama-2007} H. Kodama, Prog. Theor. Phys. Supple. 172, 11 (2008) [arXiv: 0711.4184]


\bibitem{KKZ2}
  H.~Kodama, R.~A.~Konoplya and A.~Zhidenko,
  arXiv:0904.2154 [gr-qc].


\bibitem{MS} K.~Murata and J.~Soda,
{\it Class. Quant. Grav.} {\bf 25}, 035006 (2008) [arXiv:0710.0221].

\bibitem{KIS-2000} H. Kodama, A. Ishibashi and O. Seto, Phys. Rev. D {\bf 62},
064022 (2000).

\bibitem{Kodama-Aegean} H. Kodama, Lect. Notes Phys. {\bf 769}, 427 (2009).

\bibitem{KI-2003} H. Kodama and A. Ishibashi, Prog. Theor. Phys. {\bf 110},
701 (2003).

\bibitem{MP} R.~C.~Myers and M.~J.~Perry,
{\it Annals Phys.}\  {\bf 172}, 304 (1986).

\bibitem{KKZ-2009}
  H.~Kodama, R.~A.~Konoplya and A.~Zhidenko,
  Phys.\ Rev.\  D {\bf 79}, 044003 (2009)
  [arXiv:0812.0445 [hep-th]].

\bibitem{KZ2}
  R.~A.~Konoplya and A.~Zhidenko,
  Phys.\ Rev.\  D {\bf 78}, 104017 (2008)
  [arXiv:0809.2048 [hep-th]].



\bibitem{IUM-2003} D. Ida, Y. Uchida and Y. Morisawa, Phys. Rev. D {\bf 67},
084019 (2003).

\bibitem{Abramowitz} M. Abramowitz and I. Stegun, {\it Handbook of Mathematical
Functions} (Academic, New York, 1996).

\bibitem{Berti-2006}
E.~Berti, V.~Cardoso and M.~Casals,
Phys.\ Rev.\  D {\bf 73} (2006) 024013 [Erratum-ibid.\  D {\bf 73}
(2006) 109902];

\bibitem{Cardoso-2005}
V.~Cardoso, G.~Siopsis and S.~Yoshida,
Phys.\ Rev.\  D {\bf 71}, 024019 (2005).

\bibitem{Suzuki:1998vy}
  H.~Suzuki, E.~Takasugi and H.~Umetsu,
  Prog.\ Theor.\ Phys.\  {\bf 100} (1998) 491
  arXiv:gr-qc/9805064;\\
  R.~A.~Konoplya and A.~Zhidenko,
  Phys.\ Rev.\  D {\bf 76}, 084018 (2007)
  arXiv:0707.1890 [hep-th].

\bibitem{Leaver-1985} E.W. Leaver, Proc. R. Soc. A {\bf 402}, 285 (1985).

\bibitem{Rubin.M&Ordonez1984}
M.A. Rubin and C.R. Ord\'onez: {\em J. Math. Phys.} {\bf 25}, 2888 (1984).

\bibitem{Doukas}
J.~Doukas, H.~T.~Cho, A.~S.~Cornell and W.~Naylor,
  arXiv:0906.1515 [hep-th].


\end{thebibliography}
\end{document}